\documentclass{article}

\usepackage{arxiv}

\usepackage[utf8]{inputenc} 
\usepackage[T1]{fontenc}    
\usepackage{hyperref}       
\usepackage{url}            
\usepackage{booktabs}       
\usepackage{amsfonts}       
\usepackage{nicefrac}       
\usepackage{microtype}      
\usepackage{lipsum}		
\usepackage{graphicx}     
\usepackage{xspace}     
\usepackage{booktabs}     
\usepackage{enumerate}
\usepackage{amsmath, amssymb}
\usepackage[vlined,linesnumbered,ruled,noend]{algorithm2e}
\usepackage{subfig}
\usepackage{tabularx}
\usepackage{balance}
\usepackage{multirow}
\usepackage{amssymb}
\usepackage{pifont}
\usepackage{adjustbox}
\usepackage{array}
\usepackage{xcolor,colortbl}
\usepackage{rotating}

\usepackage{float}
\usepackage{balance}
\usepackage{amsmath}
\usepackage{mathtools}

\newcommand{\hpw}{HPW\xspace}
\newcommand{\hpws}{HPWs\xspace}

\newcommand{\hpwlongs}{Hyper-partisan websites\xspace}

\newcommand{\visitedSites}{556\xspace}

\newcommand{\leftPrices}{\$0.561\xspace}
\newcommand{\rightPrices}{\$0.667\xspace}
\newcommand{\diffPrices}{5$\times$\xspace}
\newcommand{\alexa}{\texttt{alexa.com}}

\title{Stop tracking me Bro! Differential Tracking of User Demographics on Hyper-Partisan Websites}

\author{
      Pushkal Agarwal\\
      King's college London\\
      London, UK\\
      \texttt{pushkal.agarwal@kcl.ac.uk} \\
       \And
    Sagar Joglekar\\
    King's college London\\
    London, UK\\
    \texttt{sagar.joglekar@kcl.ac.uk} \\
     \And
    Panagiotis Papadopoulos\\
    Brave Software Inc.\\
    London, UK\\
    \texttt{ppapadopoulos@brave.com} \\
     \And
    Nishanth Sastry\\
    King's college London\\
    London, UK\\
    \texttt{nishanth.sastry@kcl.ac.uk} \\
     \And
    Nicolas Kourtellis\\
    Telefonica Research\\
    Barcelona, Spain\\
    \texttt{nicolas.kourtellis@telefonica.com} \\
}

\begin{document}
\maketitle

\begin{abstract}
Websites with hyper-partisan, left or right-leaning focus offer content that is typically biased towards the expectations of their target audience.
Such content often polarizes users, who are repeatedly primed to specific (extreme) content, usually reflecting hard party lines on political and socio-economic topics.
Though this polarization has been extensively studied with respect to content, it is still unknown how it associates with the online tracking experienced by browsing users, especially when they exhibit certain demographic characteristics.
For example, it is unclear how such websites enable the ad-ecosystem to track users based on their gender or age.
In this paper, we take a first step to shed light and measure such potential differences in tracking imposed on users when visiting specific party-line's websites.
For this, we design and deploy a methodology to systematically probe such websites and measure differences in user tracking.
This methodology allows us to create user personas with specific attributes like gender and age and automate their browsing behavior in a consistent and repeatable manner.
Thus, we systematically study how personas are being tracked by these websites and their third parties, especially if they exhibit particular demographic properties.
Overall, we test 9 personas on 556 hyper-partisan websites and find that right-leaning websites tend to track users more intensely than left-leaning, depending on user demographics, using both cookies and cookie synchronization methods and leading to more costly delivered ads.
\end{abstract}


\section{Introduction}
\label{sec:introduction}

In the era of mass Web monitoring, users are being tracked and their behavioral data collected and used, typically for \emph{ad-targeting purposes}.
Recent research has explored the impact of ad-tracking on user privacy~\cite{costOfAds,panpap_www2019}, and showed the scale and sophistication of the ad-ecosystem, and ad-pricing used during the real-time bidding protocol (RTB)~\cite{papadopoulos2017if} or header bidding protocol~\cite{pachilakis2019headerbidding} to serve ad-slots to webpages.
In fact, a deeper examination of the ad-ecosystem reveals a set of complex techniques such as synchronization of cookies across websites~\cite{panpap_www2019,castelluccia2014selling,acar2014web} and fingerprinting of user devices~\cite{Nikiforakis:2013:CME:2497621.2498133,devFingerprinting}, in order to perfect the user profiles being targeted.
Indeed, a highly precise profile allows ad-platforms to effectively match ads with target audiences.
However, the side effect of these highly sophisticated profiling techniques is a massive breach of user's privacy.
Without any consent, a profiled user is unknowingly disclosing her interests and (dis)likes to the benefit of ad-platforms.

Recently, such user profiling techniques have also been used \emph{for targeted opinion shaping} such as Cambridge Analytica's U.S. presidential campaigns for Donald Trump~\cite{CATrump} and Ted Cruz~\cite{CACruz}, UK-based campaigns for the Brexit referendum~\cite{CABrexit}, and Russia's social media operations during the 2016 U.S. presidential election~\cite{RussiaFB}. 
All these campaigns have one common motif: use of fake and partisan disinformation campaigns.
In fact, a slew of websites hosting such content have emerged since the start of Trump's campaign~\cite{bhatt2018illuminating}.
These pages allow fake and partisan news to be shared unchecked on the social media, garnering ever increasing partisan audiences. Some examples are Infowars, Breitbart and Fox News, for the right, and Occupy Democrats, Bipartisan Report and MSNBC for the left.

At such a juncture, we are compelled to ask: \emph{Do these fake and Hyper-partisan news websites (\hpws) -- which have been shown to have highly selective audiences -- exhibit any particular differential behavior when it comes to tracking their online visitors?}
i.e., are websites on the left (right), differentially tracking the left (right)-leaning visitors more than the right (left) leaning ones?

To answer this multi-faceted question, we first establish a methodology for understanding how \hpws and their embedded third parties track different user demographics.
We create 9 carefully crafted personas representing different genders and age groups.
We load browsers with these personas and visit a list of \visitedSites verified \hpws~\cite{bhatt2018illuminating}, to observe differences in the way these personas are being tracked via different types of cookies placed by \hpws and the third party ad-ecosystem.

We search for large-scale patterns by co-clustering both the personas and websites visited, using non-negative matrix factorization~\cite{lee1999learning,lahoti2018joint,Zhu:2007:CCL:1277741.1277825}.
We examine the extent to which the ad-ecosystem performs unbalanced tracking and cookie synchronization~\cite{acar2014web,panpap_www2019} on users of particular demographics, when a persona visits the right- or left-leaning \hpws.
Finally, we intercept and analyze RTB-based ads placed when visiting \hpws with particular personas, and establish differences between tracking practices on right and left \hpws, as well as differences in the monetary value of placed ads.

With this study, we make the following main contributions:

\begin{enumerate}
    \item We design the first to our knowledge methodology to detect possible unbalanced tracking performed from \hpws and their collaborating third parties of users that belong to different demographics.
    
    \item We extend an existing crawling tool with our methodology, and open source our framework\footnote{Data and code available from \url{https://tiny.cc/partisan-tracking/}} in order to enable fellow researchers, policy makers or even end-users to audit websites on how they personalize tracking technology based on the visitor's web profile.
    
    \item 
    Our results show that in \hpws, \textit{advertisers set the majority of cookies} on users' browsers.
    \textit{These advertisers are among the top in the overall ad-ecosystem}, and are over-represented on \hpws compared to their presence on the general web. In fact, some of the top trackers are twice as prevalent on \hpws as on the general web.
    Also, having an established persona from a particular demographic (with cookies obtained from visiting stereotypical websites for users of that demographic) results in \textit{up to 15\% more cookies} stored than for a baseline with no set persona.
    Furthermore, popular or highly-ranked \hpws track users more intensely than lower-ranked websites.
    More importantly, our results show that right-leaning websites, in general, track users with \textit{up to 25\% more cookies} than left-leaning websites. 
    They also enable more data sharing with advertisers and trackers by facilitating \textit{up to 50\% more cookie synchronizations} between these third parties, than in left-leaning websites.
    This intense tracking leads \textit{up to 5x higher cost ads} than left-leaning websites.
    We discuss implications of our work for online users' privacy.
    
\end{enumerate}

\section{Overview of Methodology}\label{sec:method}

\subsection{Third party tracking on the Web}

Websites nowadays consist of several components that may originate from many different domains other than the one the user visited.
These components provide functionality like widgets, analytics, targeted ads, and recommendations.
In order to provide as much personalized content as possible, these third parties keep track of user's personal information (e.g., geolocation, browsing device, gender) and preferences (e.g., purchases, searches, etc.) locally in a cookie placed in the user's browser, and in their servers' database using the same cookie ID.

Although several policies (e.g., Intelligent Tracking Prevention~\cite{itp} and Same Origin Policy~\cite{sop}) have been proposed to mitigate the privacy intrusion from such pervasive tracking, there are several sophisticated techniques that allow trackers to bypass such mechanism (e.g., Cookie Synchronization~\cite{panpap_www2019,bashir2016}, Web Beacons~\cite{harding2001cookies}, Cross-Device Tracking~\cite{solomos2019cdt}, etc.) and re-identify a user across different websites and create detailed profiles about her preferences and interests (e.g., sexual preferences, political beliefs, etc.)~\cite{jernigan2009gaydar}.
Then, this data can be sold to anyone interested~\cite{toysmart}, or handed over to agencies~\cite{nsaCookies}.

\begin{figure}
    \centering
    \includegraphics[width=1\columnwidth]{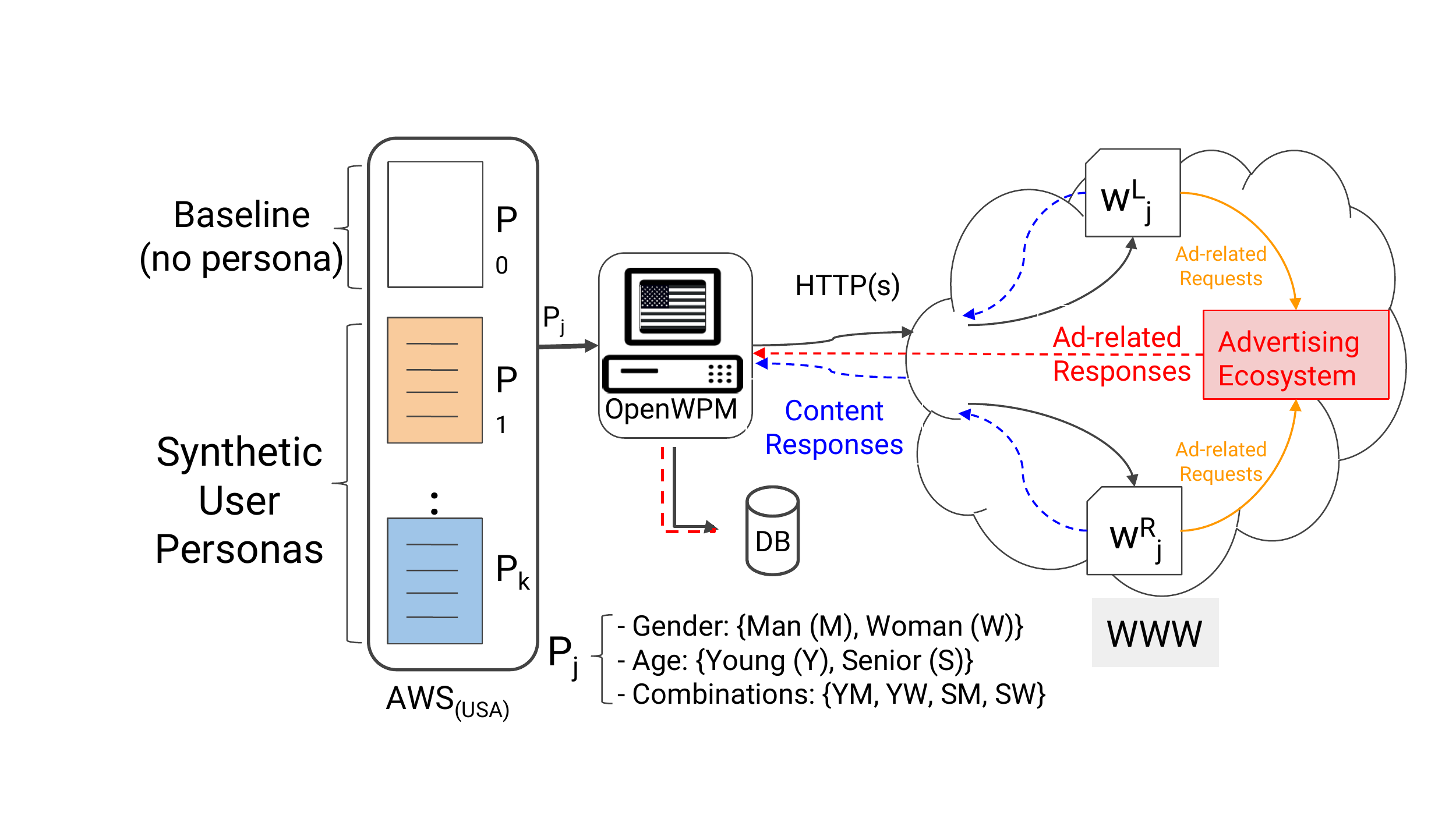}
    \caption{Crawling methodology and framework for measuring tracking of different personas by hyper-partisan websites visited, and third parties embedded in them.}
    \label{fig:machinery}
\end{figure}

\subsection{Overview of Crawling Methodology}

A general illustration of our methodology and implemented framework is shown in Figure~\ref{fig:machinery}.
To obtain consistent user behavior that allows repeatable and systematic web measurements, the proposed framework leverages a methodology with carefully curated user personas that have a history built on Web traffic that could be expected from users of specific demographic groups.
Using each of these defined personas, we then visit \hpw, which have been carefully categorized (manually) as left- or right-leaning by domain experts (journalists and fact checkers at Buzzfeed News~\cite{Buzzfeed,BuzzfeedGithub}).
Then, we log and monitor first and third party (e.g., ad-ecosystem) tracking performed during these visits, for later analysis and comparison between personas and baseline or null users.
We summarize various terms used in the paper in Table~\ref{table:notations}.

\begin{table}
    \centering
    \caption{Terms and notations used in our methodology.}
    \begin{tabular}{p{4.4cm}p{12cm}} 
        \toprule
        Terms&Notations\\
        \midrule
        General persona			& $P_j$, where $P_0$ is baseline (i.e., does not have any profile or user history loaded)\\
        Simple persona				& $P_{1\leq j \leq 4}$ are 1-feature personas\\
        Rich persona				& $P_{5\leq j \leq 8}$ are 2-feature, ``rich history'' personas \\
        \midrule
        Hyper-partisan websites (Left)		& $W^{L_i}$ are left-leaning websites; $1\leq i \leq 164$ \\
        Hyper-partisan websites (Right)	& $W^{R_i}$ are right-leaning websites; $165 \leq i \leq 556$ \\
        \midrule
        Tracking Domains (Left)		& $D^{L_i}$ are domains in all left-leaning websites; $1\leq i \leq 164$ \\
        Tracking Domains (Right)		& $D^{R_i}$ are domains in all right-leaning websites; $165 \leq i \leq 556$ \\
        \midrule
        Crawl (SQLite Database)		& $C^{P_j}_n$ represents crawl database file for persona $P_j$, $0\leq j \leq 8$ and $1\leq n \leq \{5, 15\} $ runs of the same setup depending if it is a simple persona or a rich persona.\\
        \bottomrule
    \end{tabular}
    \label{table:notations}
\end{table}

\subsection{Incremental Persona Building}\label{sec:personaBuild}

Our goal is to visit different websites with the ``persona'' of a user from a particular demographic group, based on age or gender.
We build personas by visiting top Alexa ranking websites for different demographics.
The intuition is that the third party ecosystem that enables tracking and personalized targeting~\cite{carrascosa2015,programmaticAd,ballard2016campaigning} builds up a persona or profile of a user based on the type of websites they visit.
Thus, by visiting websites which are highly popular in a particular demographic, we establish a persona of that demographic.
Previous works have inferred that there are certain demographic and geographical features which are highly associated with a person's partisan lean~\cite{bhatt2018illuminating}, including age, gender and location (whether from a right- or left-leaning US state).
Therefore, we build our personas based on these characteristics, focusing on age and gender, and training from different locations in the USA.

\begin{figure}[t]
    \centering
    \includegraphics[width=0.8\columnwidth]{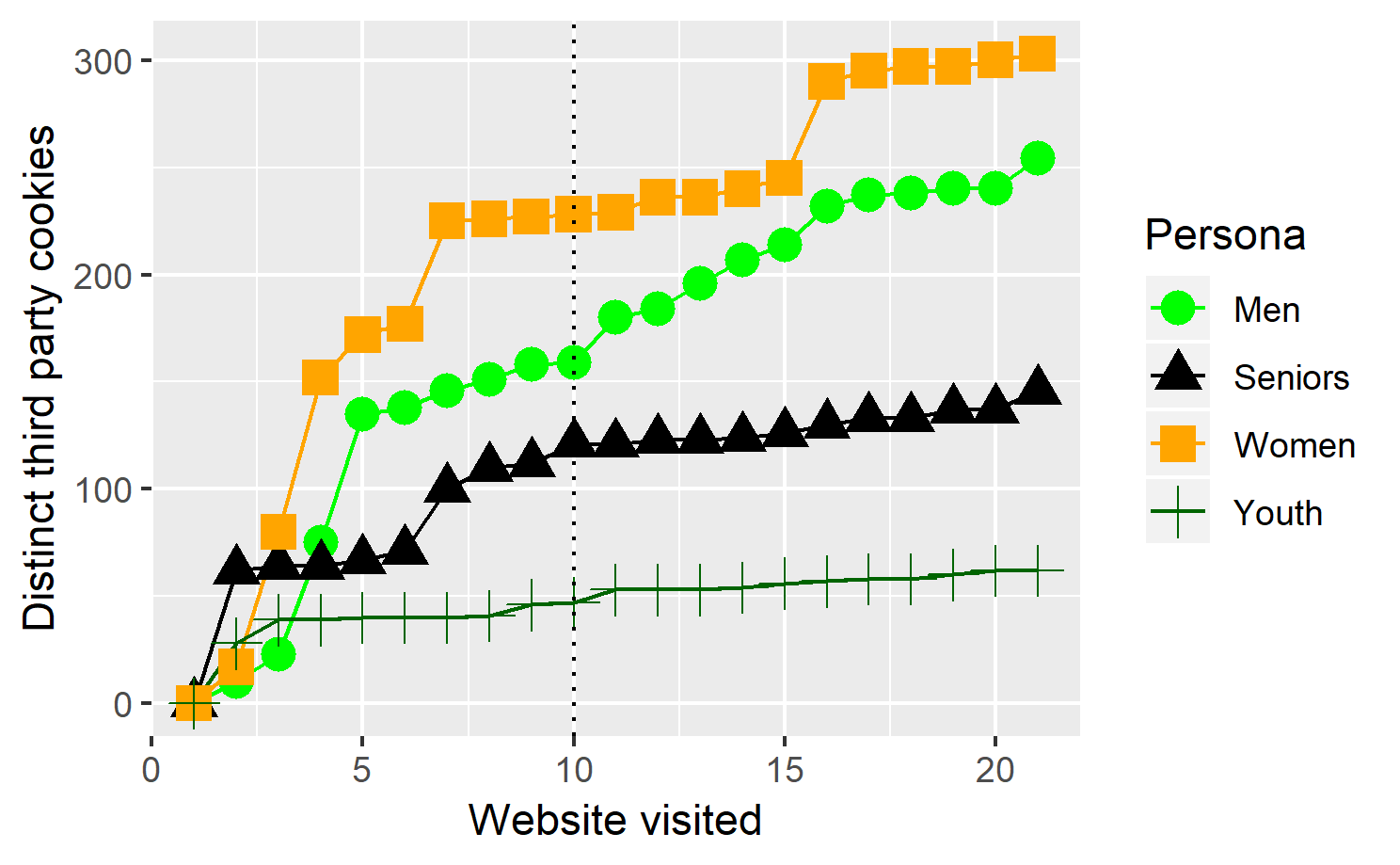}
    \caption{Numbers of distinct third parties observed after visiting a set of top \alexa\ websites for building persona per category (man, woman, young, senior).}
    \label{fig:distinctCookies}
\end{figure}

We use \alexa\ which gives the list of websites that are most popular with different demographics\footnote{Under \url{https://www.alexa.com/topsites/category}}, as captured in October 2018.
However, an important question is: \emph{How many of the websites should be visited, before a user profile becomes ``stable''?}
To check this, we visit the top websites in each category in a random order\footnote{We have verified that the actual order of visit does not affect the results}, and count the number of distinct third party cookies obtained after each new visit.
Figure~\ref{fig:distinctCookies} shows the number of third party cookies for personas with different demographics, after visiting the top websites for each category.
We observe a large increase in the number of cookies when the first 2--3 websites are visited, and then the increase in absolute numbers of distinct cookies tapers off. 
Following the average number of 30 cookies found per website in a study of 1 million Alexa-ranked websites~\cite{englehardt2016online}, we say that a persona has ``matured'' once it has reached $\geq$ 50 unique third party cookies.
We observe that visiting 6--7 websites is sufficient for a persona to mature, except in the case of Youth, which does not receive more than 100 cookies (possibly due to laws such as COPPA~\cite{coppa} which safeguard against collecting personal information on very young users) even after visiting 20 websites.
However, even in the case of the Youth demographic, we observe that the first 2--3 website visits already expose the user to the majority of distinct third party cookies, and by the time 6--7 websites are visited, well over 50\% of the eventual total number of cookies have been collected. 

Therefore, we conservatively assume a persona with a specific demographic being ``stable'' or ``established'' by visiting the top 10 websites associated with users of that demographic, and storing the third party cookies obtained.
Also, visiting the same websites from different locations produced similar numbers of cookies.
When visiting other websites (e.g., the \hpws in \S\ref{sec:hpw}), we mimic a specific demographic by loading cookies obtained for that user group.
Similar numbers of websites have been used in other studies to establish a profile of cookies.
For example, in~\cite{lecuyer2015sunlight} they used the top 10 websites of each of 16 categories for analyzing ad content. 

Note that we also use compound personas with more than one demographic characteristic (e.g., young man, or senior woman).
To establish compound personas, we visit and store cookies from the top 10 websites from each demographic (e.g., 10 of the top websites for seniors and 10 for women collectively establish the ``Senior Woman'' persona), in a random order.
We encode each persona with the initial letter of the feature incorporated in the persona.
For example `SW' signifies a persona representing \emph{Senior-Women}.
Similarly a YM persona represents the persona of a Young-Man.
The personas used are listed in Table~\ref{table:Alexa}.

We visit all websites in a persona-specific list as a stateful crawl that stores the user history in a browser.
After the visits are finished, we dump the browser state accumulated as an archive file per persona.
This persona state corresponds to stored cookies, HTTP calls done, location, etc.
Each archived file can then be used to bootstrap a browser with the said persona, ensuring our measurements per persona for each visit of a \hpw are bootstrapped with the same browser persona state.
Next, we discuss more details on automation and browser settings of the crawling approach.

\begin{table}
    \centering
    \caption{Examples of websites from each category of \alexa\\for creating personas with specific demographics.}
    {\small 
        \begin{tabular}{p{1.8cm}p{5.7cm}} 
            \toprule
            Demographic	&	Sample websites	\\
            \midrule
            Youth (Y)		&	(Student, Kidzworld).com, Voicesofyouth.org \\
            Senior (S)		&	Aarp.org, (Medicare, Cms).gov \\
            \midrule
            Woman (W)	&	(Cosmopolitan, Womansday, Sheknows).com \\
            Man (M)		&	(Menshealth, Mensjournal, Esquire).com \\
            \bottomrule
        \end{tabular}
    }
    \label{table:Alexa}
\end{table}

\subsection{Crawling Engine of Framework}\label{sec:crawling}

We make use of \emph{OpenWPM}\cite{openwpm-code}, a popular tool for measurements and automating web browsers, which is developed by Princeton University under the Web Transparency \& Accountability Project~\cite{englehardt2016online}.
Using OpenWPM, we create user personas and bootstrap parallel sessions of \emph{Firefox} browsers to visit \hpws with each persona.
In order to be able to compare browsing experience of a persona with a baseline, we load null personas (i.e., a user with no website visit history) to the browser and re-execute the website auditing.
Also, using the same tool, we collect cookies set both by javascript calls (i.e., transactional cookies) and HTTP calls (regular cookies set in the browser, logged as Profile cookies).
Transactional cookies include all logs of cookies set in the browser, including the deleted instances of cookies.
In our analysis, we consider both types and refer to them as \emph{Cookies}.

We kept most of the default settings of OpenWPM's browser instance and updated a few for our use.
The default settings include loading a Firefox browser (only this is supported by the tool), having always enabled third party (tp\_cookie) tracking, keeping blockers like donottrack, disconnect and ghostery, https-everywhere, adblock-plus, ublock-origin and tracking protection as false, so that nothing is blocked.
The updates include enabling the http\_instrument which logs HTTP responses, requests and redirects, using the selenium headless browser to perform crawling and, setting js\_instrument, cookie\_instrument and save\_javascript to true to store javascript cookies table, regular (profile) cookies table and all javascript snippets loaded and executed by website respectively.

A \hpw crawl with a loaded persona is stateless, i.e., each \hpw website visit is independent.
We repeated each \hpw crawl 5 times per persona and 2 times per baseline, to account for infrequent, but unavoidable network errors.
In later sections, we label a persona visiting a left- or right- \hpw, using `L' and `R' respectively, and append this with the persona's acronym.
For example, `L:Y' is for a \emph{Youth} persona visiting \emph{left} \hpws; similarly `R:SM' is for a \emph{Senior-Man} persona visiting \emph{right} \hpws.

Since each browser instance is independent, multiple browser instances can be initiated on a crawling server (depending on its available resources).
This parallelization allows for multiple, simultaneous crawls of personas and baselines, scaling our auditing to many websites at the same time from one location.
This also allows us to capture the same tracking and advertising effort from third parties, before ad-campaign budgets change.

Our crawls were executed in Nov-Dec 2018.
Our crawling engine was set up on 10 Amazon Web Services (AWS) instances (each: 1 GB memory, 1 core, 8 GB storage, Ubuntu 14.04) in the USA, for instantiating parallel crawls.
The crawling was orchestrated by OpenWPM, loading a Firefox 52.0 browser instance and visit websites using Selenium.
Each AWS instance uses a parallel and independent OpenWPM instance loaded with a different persona profile.
Each crawl gives us $\sim$500MB SQLlite database across the 10 instances.
This database stores information about the HTTP calls, cookies, visit sequence, and other meta information based on the settings described earlier.

\subsection{Datasets Collected}\label{sec:hpw}

We implemented our methodology into the framework illustrated in Figure~\ref{fig:machinery} for user tracking on 667 \hpws.
Each website is marked as left- or right-leaning ($W^L$ and $W^R$, respectively) based on the description and self attestation from the website's `about' page, or facebook page description.
Interestingly, out of 667 websites curated in 2016, 111 websites were not active in Nov-Dec 2018 when the crawls took place.
Thus, in our study we use 556 active websites: 164 left- and 392 right-leaning \hpws.
We normalize our results to account for the imbalance between the two lists.
We visit \hpws using baseline (null) and persona profiles.
We store the cookies served to each persona profile or new user and analyze their distributions, type of first or third party sending them, popularity of the \hpw involved, etc.
A summary of our dataset regarding HTTP calls and cookies set during persona building and final crawls for all personas and \hpws is given in Table~\ref{table:OpenWPM}.
The results on baseline profiles and personas are analyzed in the next two sections, respectively.

\begin{table}
    \centering
    \caption{Crawls for persona building (P:) and visits to \hpws (HPW:).
        Second column: total count of requests or responses across crawls;
        Third column: average count per website;
        Fourth: standard deviation (SD) per website; Fifth: median count per website.}
    {\small
        \begin{tabular}{p{3.5cm}p{1.5cm}p{2.5cm}p{2cm}p{2.5cm}}
            \toprule
            Response Type			&	Total	Count		&	Mean per website &SD per website&Median per website	\\
            \midrule
            P: HTTP Requests		&	53.2k			&	190 &186 &	135	\\
            P: HTTP Responses		&	49.6k			&	177 &155	&	132	\\
            P: HTTP Redirects		&	7k				&	25 &46	&	9	\\
            P: Cookies			&	36.8k			&	131 &192	&	67	\\
            \midrule
            HPW: HTTP Requests	&	8.3 M			&	159 &194 & 	98	\\
            HPW: HTTP Responses	&	8.1 M			&	144 &181	&	93	\\
            HPW: HTTP Redirects	&	2.5 M			&	65 &149	&	40	\\
            HPW: Cookies			&	3.5 M			&	60 &132	&	16	\\
            \bottomrule
        \end{tabular}
    }
    \label{table:OpenWPM}
\end{table}

\section{Basic User Tracking in HPW\lowercase{s}}\label{sec:results-baselines}

Our modus operandi is to use the previous methodology with our deployed framework to visit various \hpw, and examine differences in the tracking performed by the first and third parties included in each.
In this section, we focus on null profiles to measure basic tracking performed on new (null) users.
We conduct an investigation on the differences in tracking, by comparing left and right \hpws (Sec.~\ref{sec:left-vs-right-baselines}), by looking into the top trackers on the Web and whether they are over/under represented in left and right \hpws (Sec.~\ref{sec:top-trackers-baselines}), and finally, by studying popularity of \hpws and checking if this associates with differences in tracking (Sec.~\ref{sec:popularity-baselines}).

\subsection{Who facilitates more tracking: left or right?}\label{sec:left-vs-right-baselines}

\begin{figure}
    \centering
    \includegraphics[width=0.7\columnwidth]{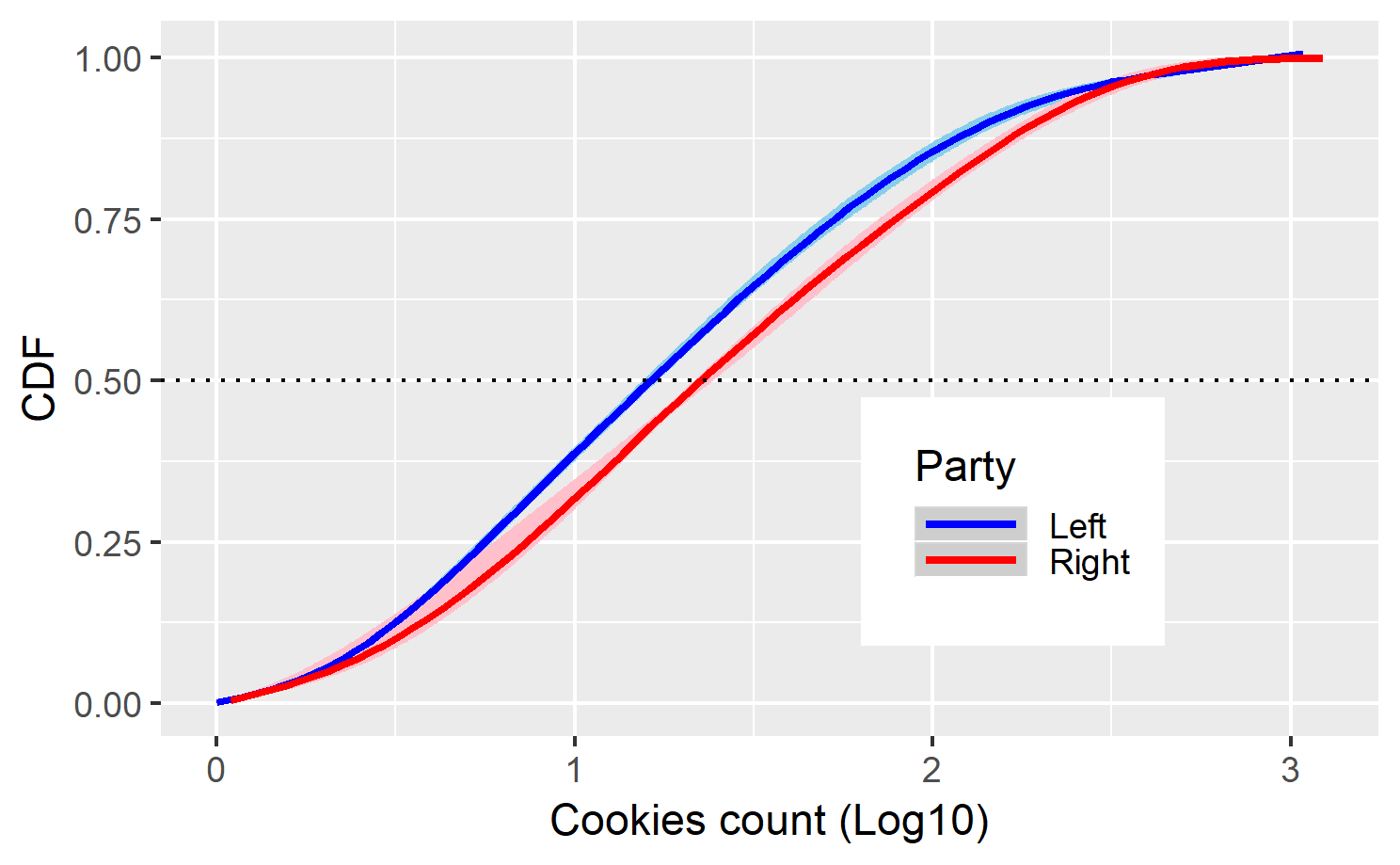}
    \caption{CDF of total number of cookies stored per website, when a baseline user visits left and right-leaning \hpws.}
    \label{fig:jsCookiesCDF}
\end{figure}

The volume of cookies in a website is a good proxy for measuring how much the website tracks a user directly, and how much it enables third parties in its page to track this user.
To understand how much of this tracking is happening on \hpws, in Figure~\ref{fig:jsCookiesCDF}, we plot the Cumulative Distribution Function (CDF) of number of cookies set on a baseline user's profile when visiting $W^L$ and $W^R$.
On average, $W^R$ set 9 more cookies than $W^L$ on baseline users.

\begin{figure}[t]
    \centering
    \includegraphics[width=0.8\columnwidth]{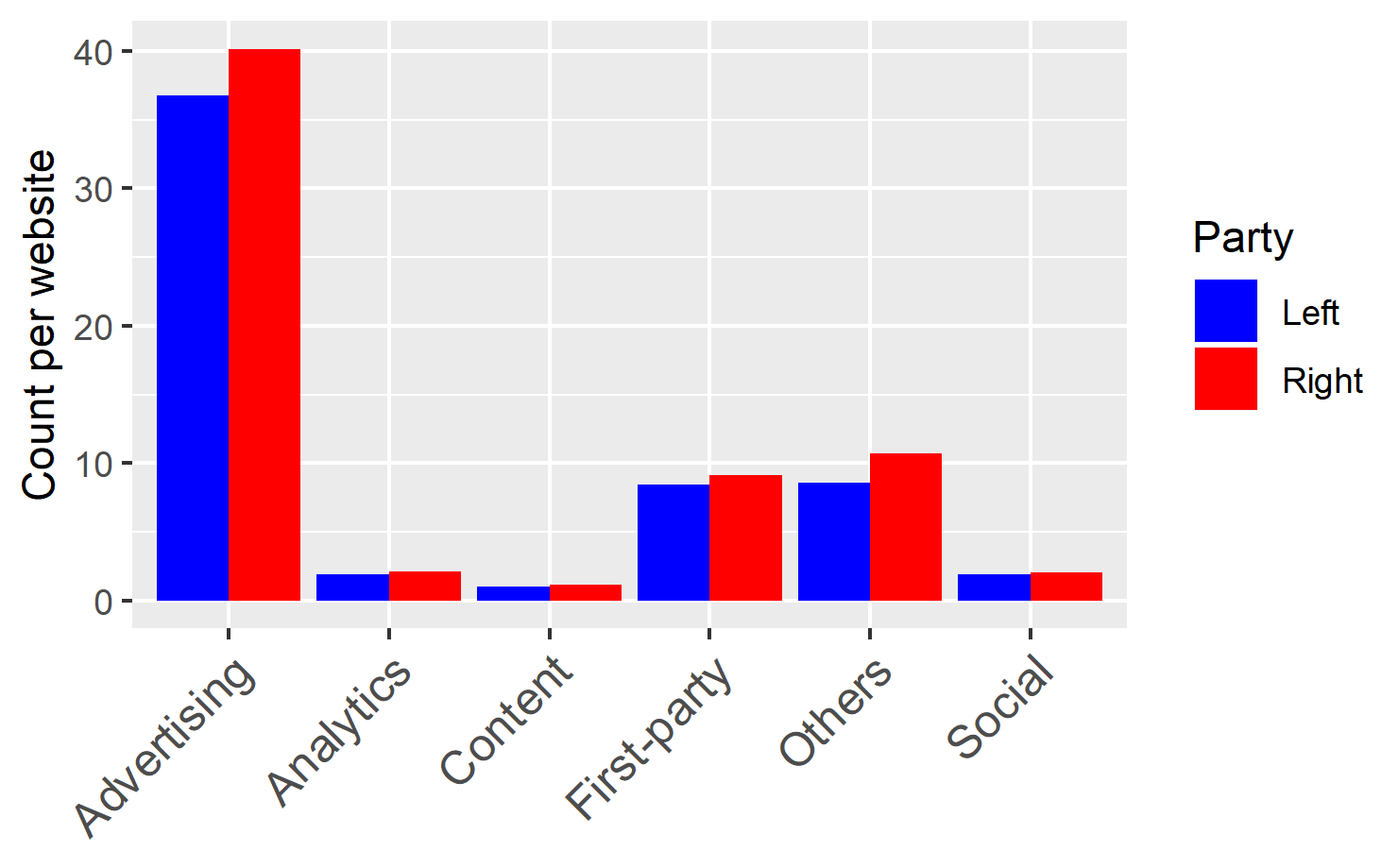}
    \caption{Average number of cookies per website, per type of domain sending them (using the \texttt{Disconnect.me} categories).}
    \label{fig:jscType}
\end{figure}

Next, we study the type of cookies set on baseline users, using a list of \texttt{Disconnect.me}~\cite{disconnectme}.
In Figure~\ref{fig:jscType}, we breakdown the cookies into six types: \emph{first-party, advertising, analytics, content, social, and other}.
Third party cookies from advertising entities significantly outnumber all other types.
Furthermore, $W^R$ place more cookies of all types, and especially for advertising, in comparison to $W^L$.

\subsection{Is this tracking more than the general Web?}\label{sec:top-trackers-baselines}
\begin{figure}
    \centering
    \includegraphics[width=0.8\columnwidth]{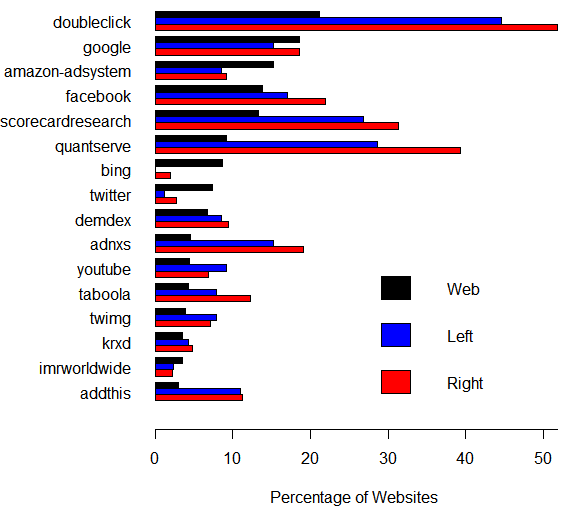}
    \caption{Percentage of websites with cookies set (x-axis) from top 16 tracking domains on the Web (y-axis), for left and right-leaning \hpws, as well as in the Web, for a baseline user.
        Domain ranking based on live list of \texttt{whotracks.me}.}
    \label{fig:toptrackers-baseline}
\end{figure}

Earlier, we showed that \hpws enable tracking of users by third parties, and that $W^R$ do so in higher intensity than $W^L$.
But who are the top trackers in these partisan websites?
And how different is the tracking they do, when a baseline user visits such \hpws?
We extract the top trackers on the Web using a live list maintained by \texttt{whotracks.me}~\cite{whotracksme} on 25/09/19, and compare this list against the trackers detected in the two groups of websites, when they are visited by a baseline user.
We find that 72\% of third parties match between the list and the ones on \hpws.
However, there are also differences among the \hpw tracking ecosystem and the Web with respect to intensity in tracking from each party.
Figure~\ref{fig:toptrackers-baseline} shows a histogram of percentage of websites that specific trackers appear in, and drop cookies on their users.
Many of these top trackers are over-represented in the \hpws examined, in comparison to the Web.
For example, \textit{Doubleclick, Scorecardresearch, Quantserve, and Adnxs}, appear in at least twice as many websites (proportionally) than in the general Web.
Furthermore, many of them appear in many more websites on the right ($W^R$) than the left ($W^L$), demonstrating an interest in tracking users visiting right-leaning \hpws.

\subsection{Is tracking associated with site popularity?}\label{sec:popularity-baselines}

\begin{figure}[t]
    \centering
    \includegraphics[width=0.8\columnwidth]{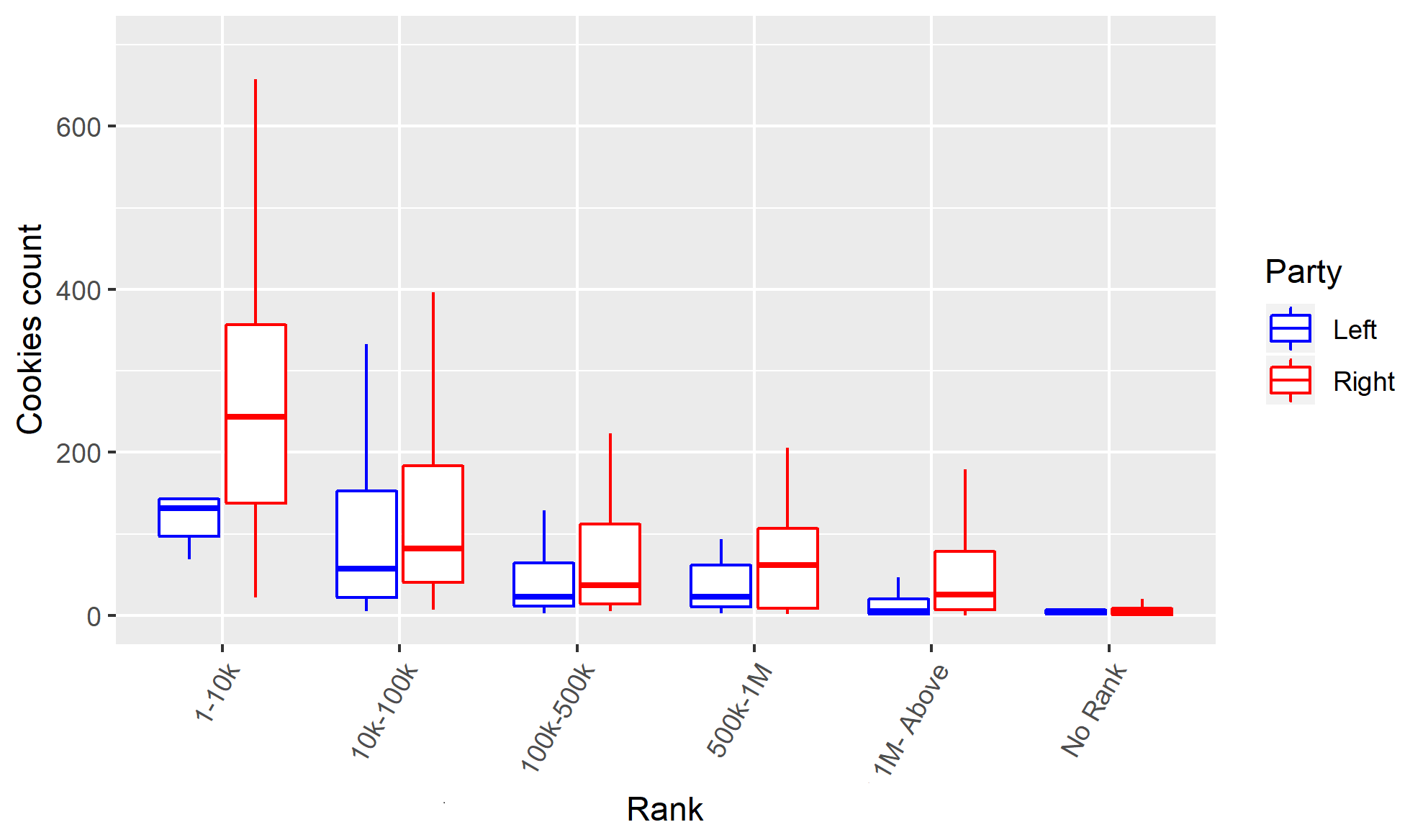}
    \caption{Cookies count for baseline users visiting $W^R$ and $W^L$, vs. the \alexa\ global ranking of these websites captured on 25/02/19.
        Top rank $W^R$ have highest median of cookie count and the median decreases as the rank increases.
        Websites with no ranking have very few cookies.} 
    \label{fig:rankBaseRange}
    \vspace{-0.5cm}
\end{figure}

\begin{figure*}[h]
    \centering
    \subfloat[HPW \& persona third party overlap] {
        \includegraphics[width=0.33\textwidth]{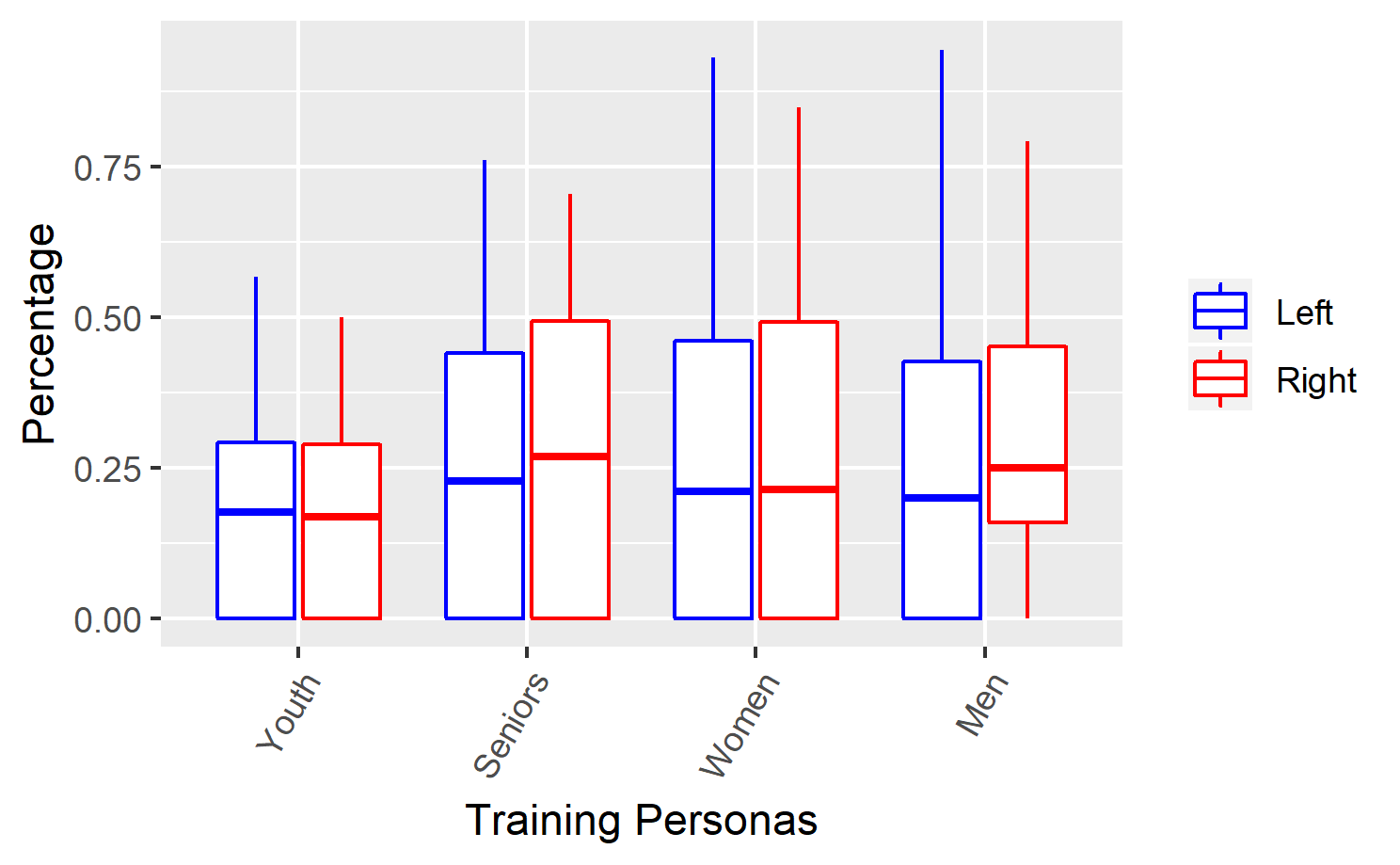}
        \label{fig:overlap}
    }
    \subfloat[KS-statistic of third party comparison] {
        \includegraphics[width=0.33\textwidth]{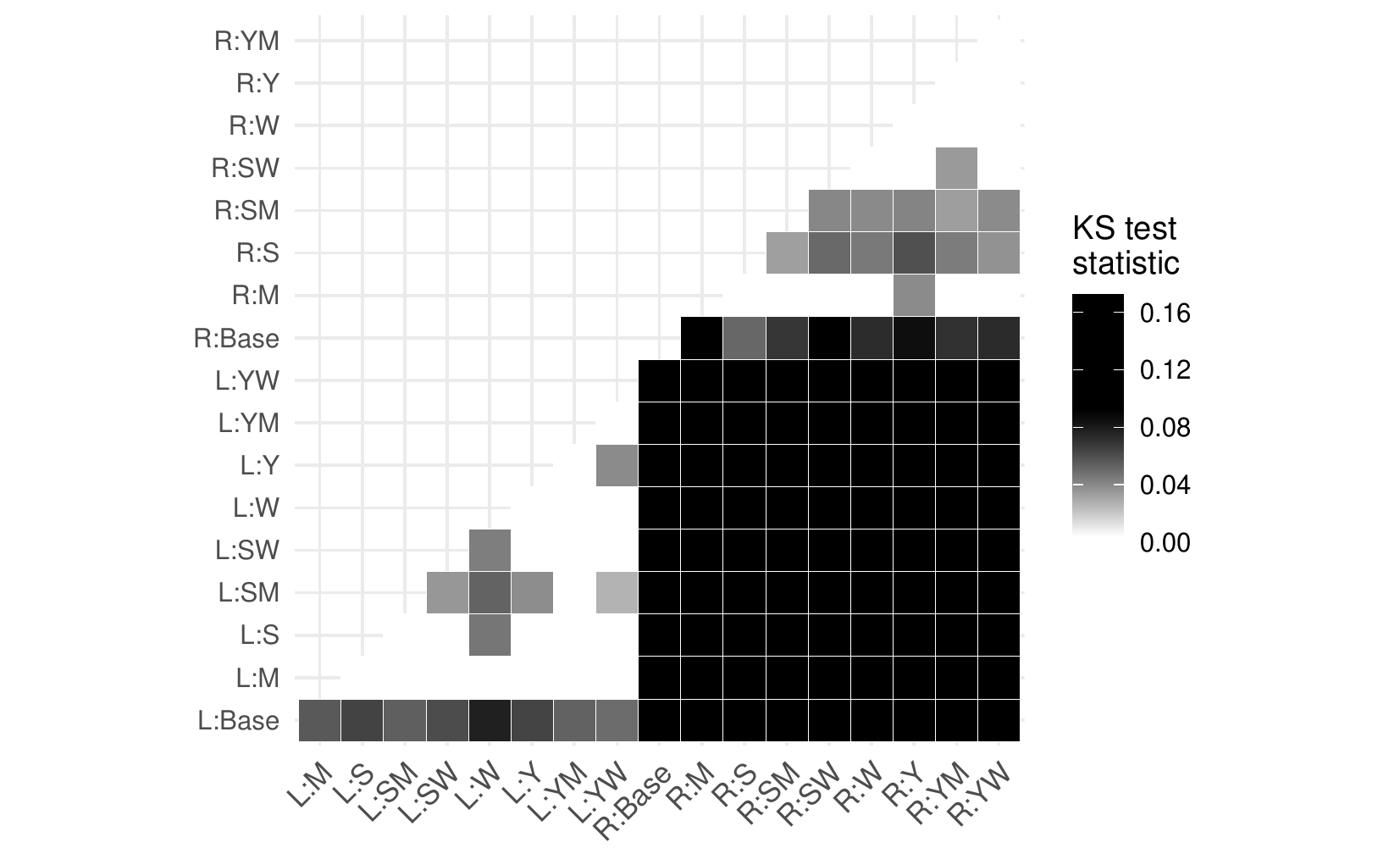}
        \label{fig:cookiesDomainKSP}
    }
    \subfloat[Difference of baseline \& personas] {
        \includegraphics[width=0.33\textwidth]{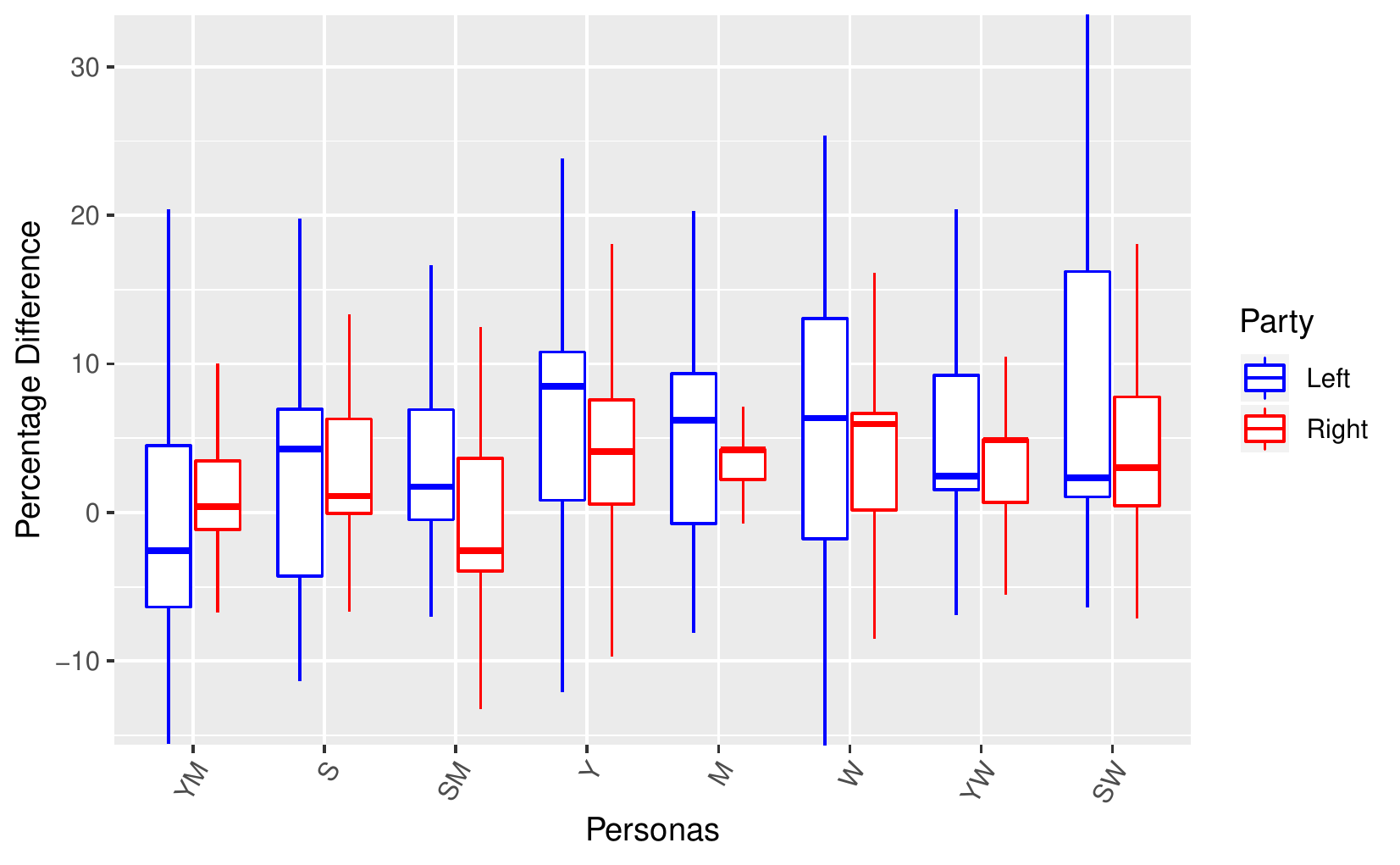}
        \label{fig:percentDiffDomains}
    }
    \caption{
        (a) Overlap between third parties dropping cookies during persona building, and when the same personas visit \hpws.
        (b) Heat-map of KS statistic test for all pairwise comparisons between distributions of numbers of unique third parties serving cookies.
        All cells with $p\geq 0.01$ are whited out; only cells with $p<0.01$ are colored.
        (c) Percentage difference between third parties serving cookies to baseline and loaded personas.
        The	x-axis is sorted on medians of all personas.}
\end{figure*}

We observed differences between $W^R$ and $W^L$ in number of cookies they drop on users, type of third parties involved, and top trackers in the Web that also appear in these websites.
Next, we investigate how the tracking intensity of \hpws associates with the popularity of each \hpw.
In Figure~\ref{fig:rankBaseRange}, we present the cookie counts of \hpws dropped on baseline users, against the respective rank range of each website based on Alexa.
Again, we find that regardless of the rank range, $W^R$ serve more cookies than $W^L$.
Interestingly, the top ranked $W^R$ (i.e., 1-10K) demonstrate the highest median count, with that count decreasing for lower ranks.

In the next section, we compute various measures to provide statistical evidence of differences in personas and their demographics vs. intensity of tracking facilitated by \hpwlongs.

\section{User Persona Tracking in HPW\lowercase{s}} \label{sec:results-personas}

The central question of this study is whether online personas exhibiting specific partisan browsing patterns are targeted in a differentiated fashion by the ads and analytics ecosystem.
We use our deployed framework to investigate this question.
Using the built personas, we first examine the overlap between third parties for distinct personas across partisan aisles.
This helps us ascertain whether our framework's personas are being treated as valid by the tracking ecosystem (Sec.~\ref{sec:overlap}).
Then, we measure statistical differences between third parties across personas and popularity of domains (Sec.~\ref{sec:more-cookies-demo}).
We also use a matrix-based clustering technique to investigate whether trackers behave preferentially across partisan lines (Sec.~\ref{sec:clustering}).
Then, we investigate the information flows between trackers, by measuring cookie synchronizations happening across webpage visits (Sec.~\ref{sec:cs}).
Finally, we evaluate whether the ad-ecosystem differentiates between our partisan personas in terms of the ad-pricing offered across different \hpws (Sec.~\ref{sec:prices}).

\subsection{Third Parties in User Personas and \hpws}
\label{sec:overlap}
To produce systematic results regarding tracking of personas, we first need to validate whether our personas are being tracked by the \hpws ad-ecosystem.
Validating this step imparts confidence on our results.
To that end, we select the personas of youth, seniors, women and men, and separately visit left and right leaning \hpws, to check the overlap of common third parties across \hpws and personas.
For this, we measure the total number of common third parties in personas and left or right \hpws.
As seen in Figure \ref{fig:overlap}, all comparisons yield upwards of 25\% overlap in domains of third party cookies.
This signifies that our personas can be actively tracked by trackers that also exist in left and right-leaning \hpws.

\subsection{Which demographic is tracked more?}\label{sec:more-cookies-demo}

Next, we compare the number of third party cookies from different domains stored for different personas (Figure~\ref{fig:cookiesDomainKSP}).
Again, we see that $W^R$ have many more third parties embedded in their pages.
They also have higher number of cookies than $W^L$, perhaps expectedly given the larger number of third parties.
This result points to a tendency from $W^R$ to monitor more intensely users across the Web with tracking technologies, with the purpose of (re)targeted advertising.

Although fewer third parties are interested in $W^L$, the top ones that are involved have a higher coverage than the left \hpws covered.
As example, most of the top third parties such as \emph{Google} and \emph{Adobe} have similar coverage of both $W^L$ and $W^R$.
However, providers such as \emph{Pubmatic} and \emph{Adsupply} have a higher coverage of and specialization in $W^L$ or $W^R$, respectively.
In cookies count, we also find extreme cases from
$W^R$ (e.g., \url{spectator.org}, \url{realclearpolitics.com}) and
$W^L$ (e.g., \url{newshounds.us}, \url{opednews.com}) which set more than a \emph{thousand} cookies.

Next, in Figure~\ref{fig:percentDiffDomains}, we compare the difference in numbers of third parties and cookies, when comparing a baseline persona with one of the curated personas.
A few personas, such as Youth Man (YM), seem to experience a slight fall in number of cookies relative to baseline.
However, most others see an increase in the number of third parties setting cookies.
From Figure~\ref{fig:percentDiffDomains}, we see that, in general, single-feature personas such as \emph{Woman, Man} and \emph{Youth}  see the highest increase in number of cookies placed, regardless of political leaning of the website.
Comparing this with Figure~\ref{fig:distinctCookies}, where \emph{Youth} hardly gets $\sim$50 cookies even after visiting 10 websites, this increase in tracking over baseline suggests unusual behavior on \hpws.
From the same figure, for $W^L$, we see the highest increase in tracking is for \emph{Youth} and for $W^R$ the highest tracking increase is for \emph{Woman}.

\begin{figure}[t]
    \centering
    \includegraphics[width=0.8\columnwidth]{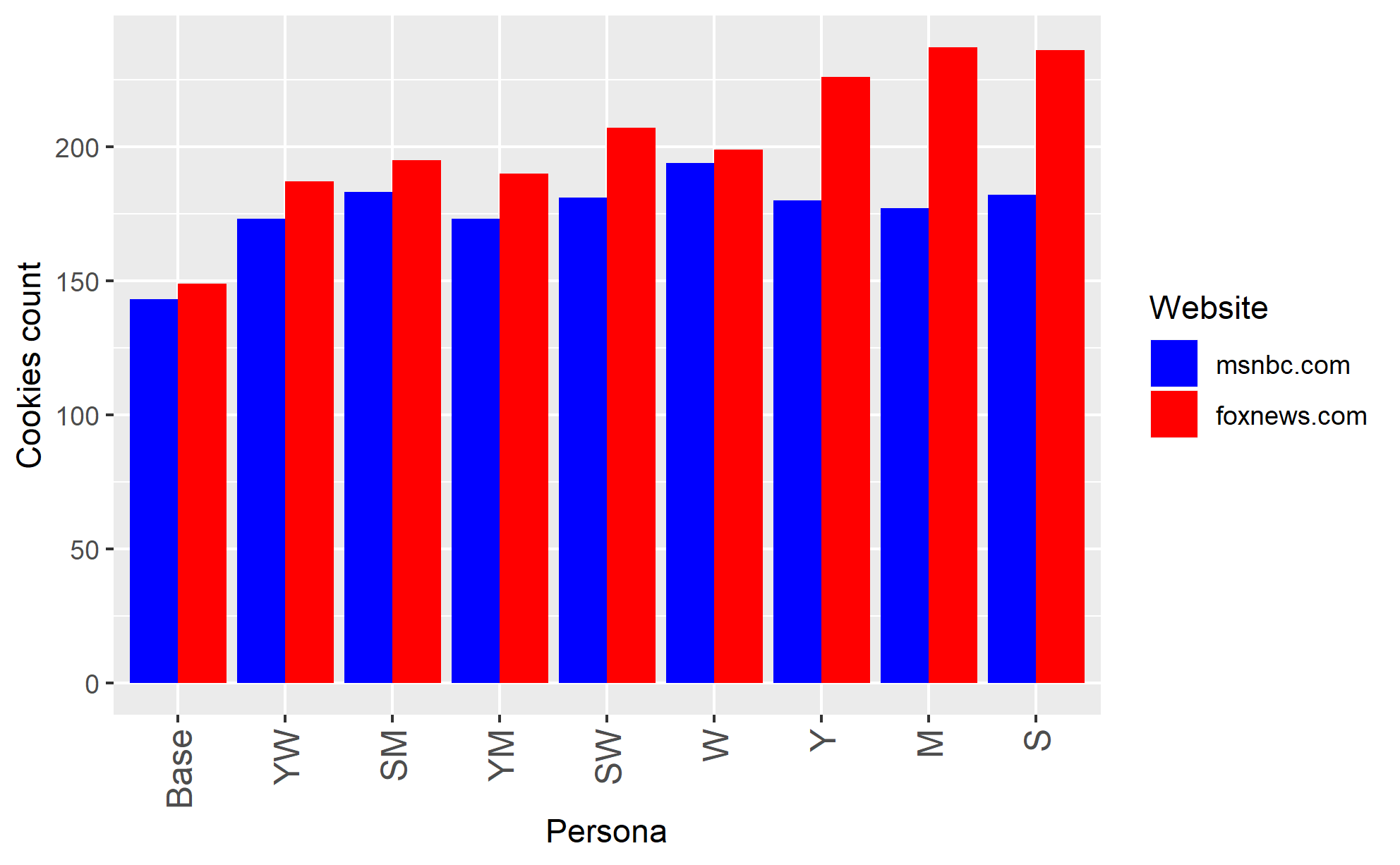}
    \caption{Variation in cookies count for foxnews.com (top ranked $W^R$) and msnbc.com (top ranked $W^L$) with various personas.
        In both websites, the cookie count is lowest for the baseline persona. X-axis is sorted by increasing count of cookies on both sides.}
    \label{fig:foxnews}
\end{figure}

\begin{figure*}[t]
    \centering
    \subfloat[B Factor]{
        \includegraphics[width=0.6\textwidth]{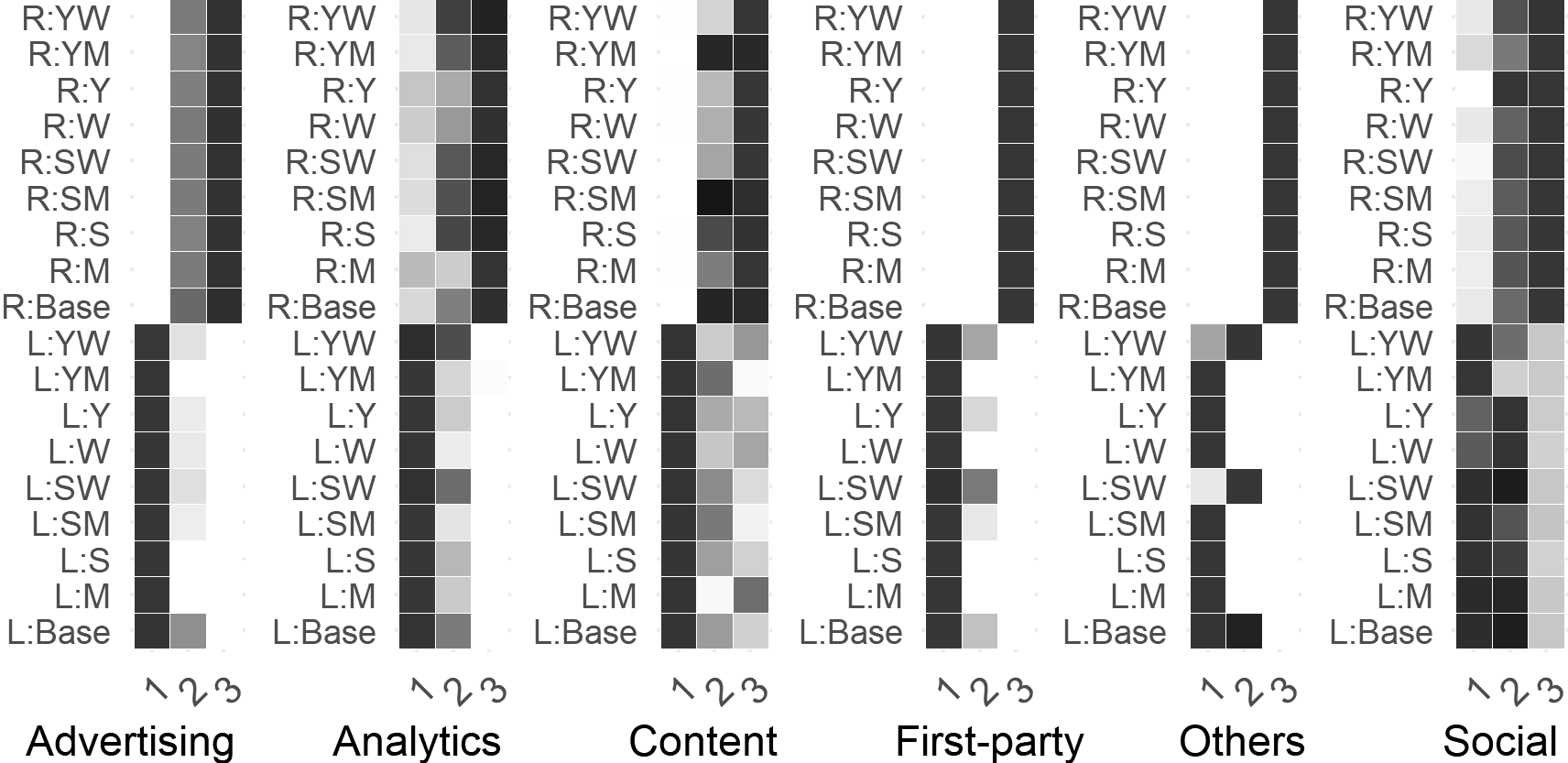}
    }
    \subfloat[C Factor]{
        \includegraphics[width=0.4\textwidth]{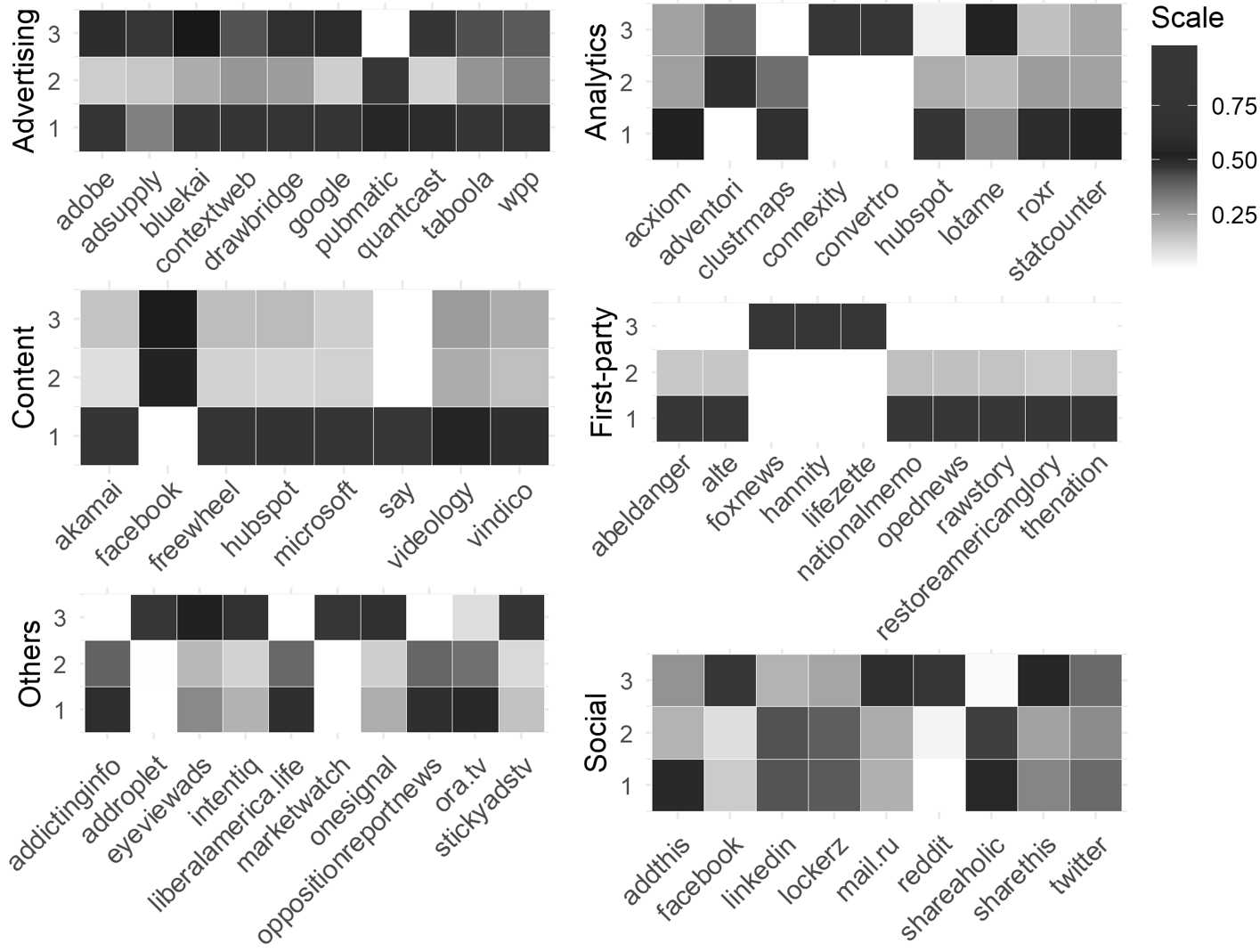}
    }	
    \caption{$B^t$ and $C^t$ factors from NMF clustering of $A^t$ for six categories and 3 clusters (i.e., $k=3$). We show the scale of all the figures in top-right corner, which is normalized from 0 to 1.}
    \label{fig:nmfW}
\end{figure*}

As a special case, we also study two popular websites: \emph{MSNBC} ($W^L$) and \emph{Fox News} ($W^R$).
Figure~\ref{fig:foxnews} shows the cookies placed by these two websites on visitors with different personas.
These examples highlight that there is a bias in the way partisan news websites change their tracking behaviors, depending on the demographic of the user visiting them.
Also note that \emph{Fox News} ($\in W^R$) serves more cookies than \emph{MSNBC} ($\in W^L$) for all personas.

\subsection{Hyper-partisan Clustering}\label{sec:clustering}

We test our framework on self declared left or right leaning websites~\cite{bhatt2018illuminating}.
Hence, we hypothesize that the trackers being deployed on these \hpws would converge onto preferentially tracking personas conducive to the prime audiences of these websites.
For example, a tracker deployed on a right leaning website would prefer deploying the tracking cookies on a right leaning persona as opposed to a left leaning persona. 

To test this hypothesis, we use Non-Negative Matrix Factorization (NMF)~\cite{lee1999learning} to cluster the websites, as well as the targeted demographics into the most viable clusters.
The members of these clusters exhibit more similar behaviors to each other, compared to members outside these clusters. 

NMF decomposes any non square matrix $A$ into two components such that $A \approx B \times C$,
where $B$ is called the factor \emph{basis} matrix with the dimensions $p$~$\times$~$k$ , and $C$ is called the \emph{coefficient} matrix with dimensions $k \times c$, for any choice of $k$ clusters.
For a given choice of $k$, the NMF algorithm tries to solve the optimization problem:
\begin{equation}
min {||A-BC||}_F^2, \ such\  that\  B\geq 0\ and\ C \geq 0
\end{equation}
where $F$ represents the \emph{Frobenius} norm.
To setup this optimization problem, we first need to compose the matrix $A$, such that it reflects the cookie profile observed by any given persona.
We create a $p$~$\times$~$c$ dimensional matrix, which we call $A$, where $p$ represents the rows corresponding to all the personas we curate, and $c$ represents the columns which correspond to the different third party.
We normalize each row, such that $A_{ij}$ now represents the percentage of cookies injected for a user of the $P_i$ persona by the $j^{th}$ domain. 

As described in Section~\ref{sec:left-vs-right-baselines}, we have distinct set of cookies for each persona viz. \emph{first-party, advertising, analytics, content, social and others}.
Hence, we create 6 distinct matrices corresponding to the 6 different types of cookies.
We further factorize each of these `persona cookie profile' matrices $A^t$, into corresponding basis $B^t$ and coefficient $C^t$ matrices where  $t\in(1,6)$ using NMF.

According to  \cite{brunet2004metagenes,hutchins2008position}, clusters should be chosen to have high cophenetic correlation and small residuals. 
After experimenting with various values of $k$($\in[2,10]$), we find that the best clustering is obtained when $k=3$, giving us the maximum cophentic correlation (0.982) and minimum residual value (9).

\begin{figure*}[thb]
    \centering
    \subfloat[Cookie synchronization] { 				
        \includegraphics[width=0.33\textwidth]{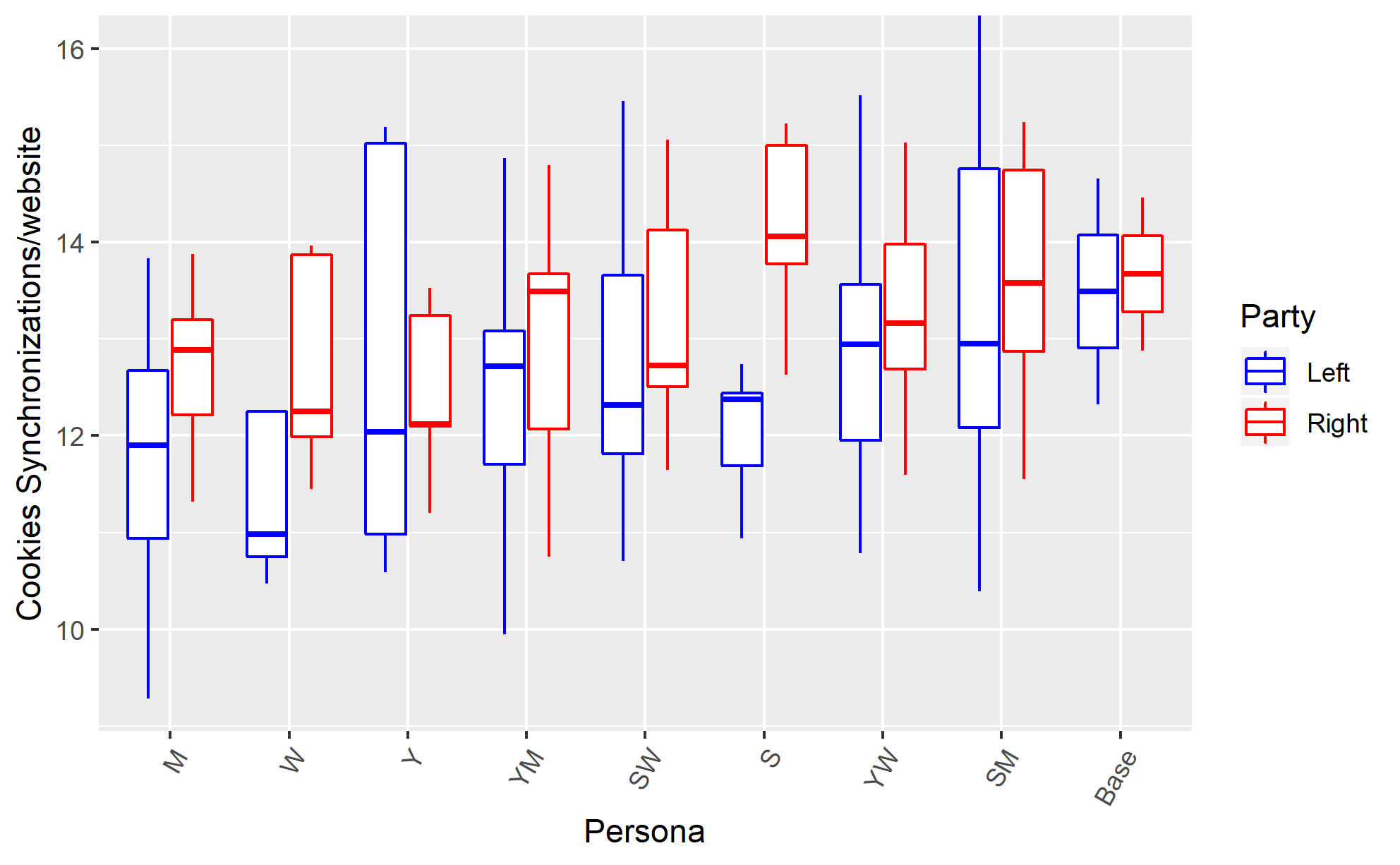}
        \label{fig:csyncBox}
    }
    \subfloat[Distribution of Prices] {
        \includegraphics[width=0.33\textwidth]{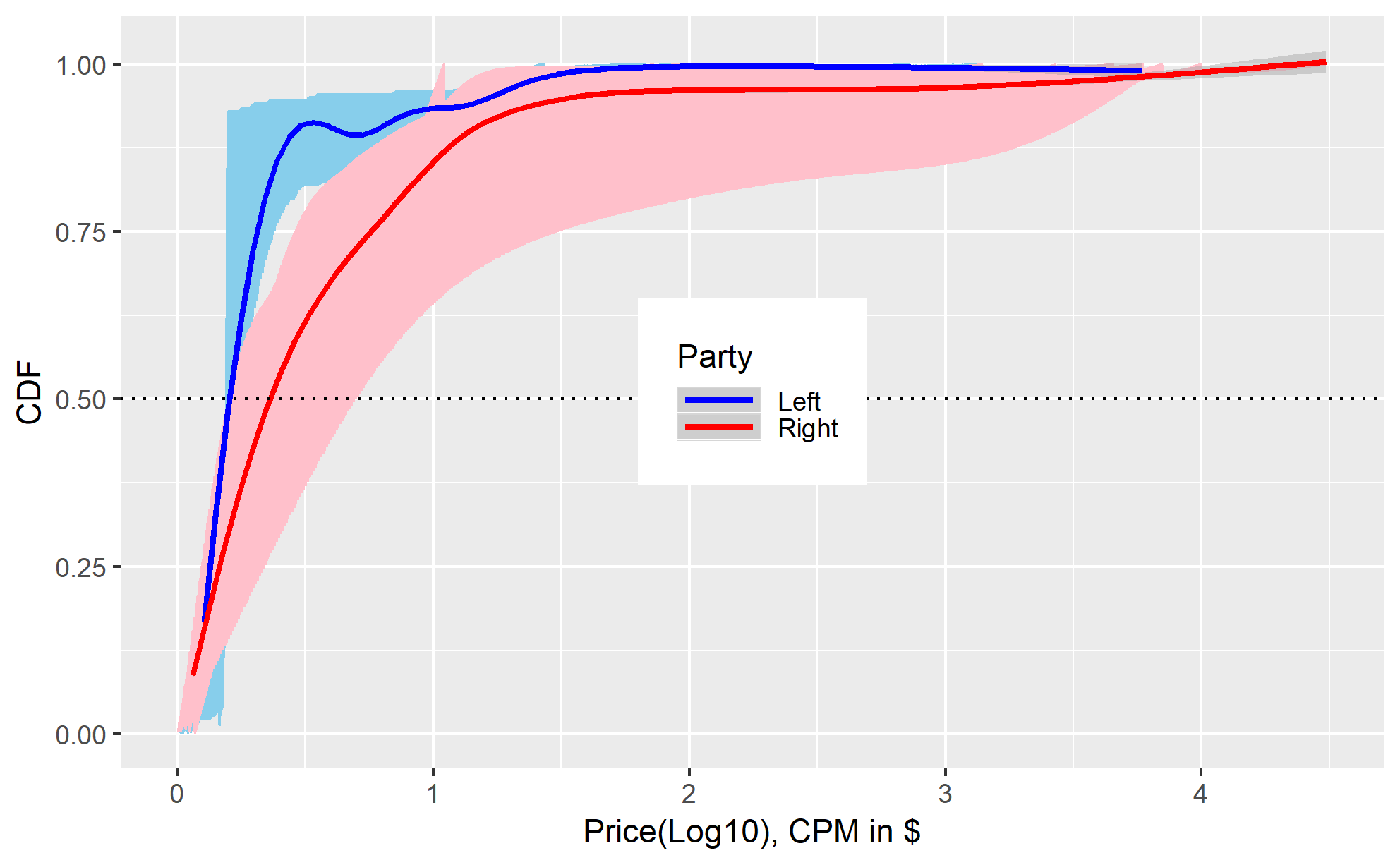}
        \label{fig:pricesCDF}
    }
    \subfloat[ KS statistic of Prices] {
        \includegraphics[width=0.33\textwidth]{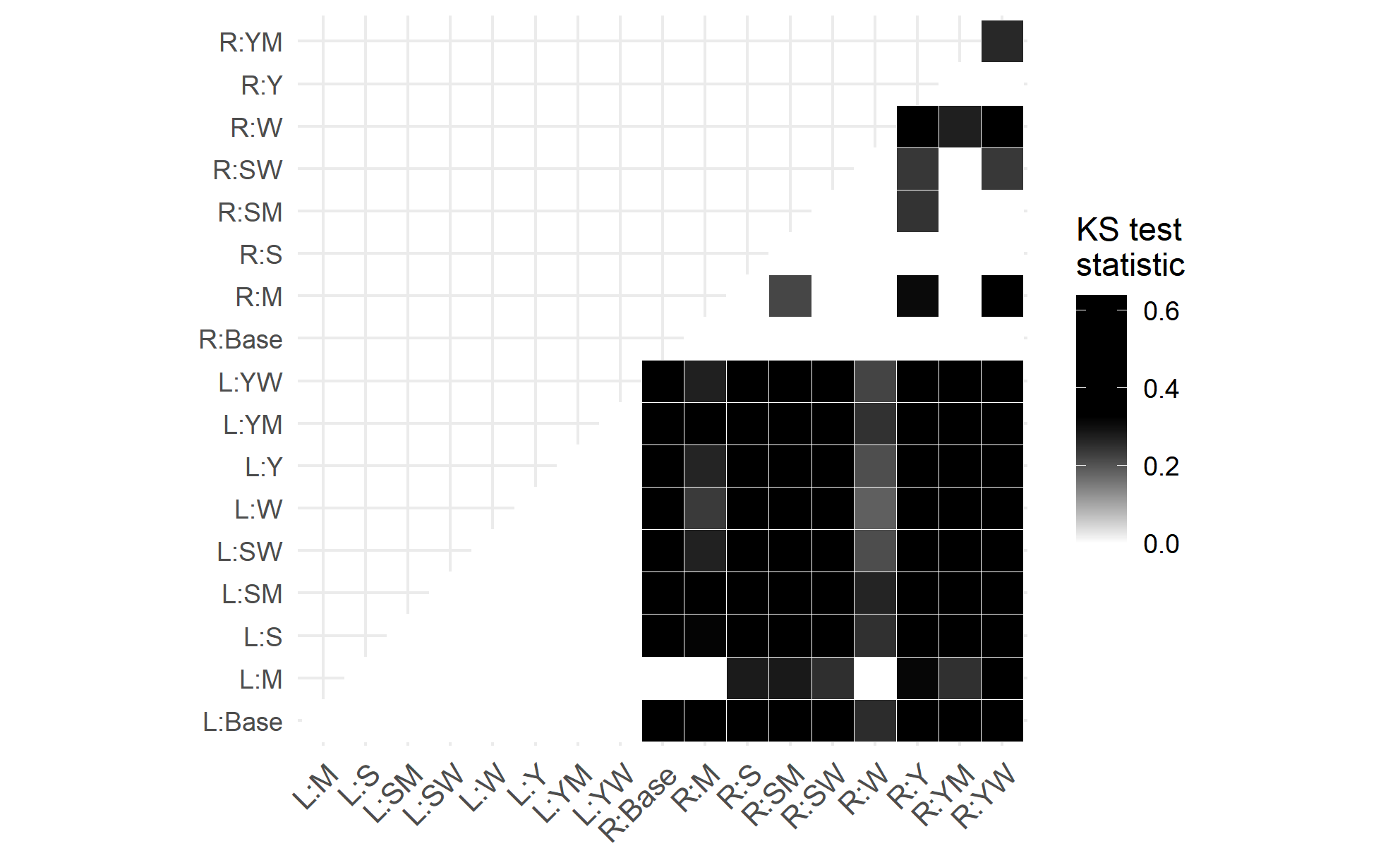}
        \label{fig:pricesKS}
    }
    \caption{(a) Cookie synchronizations per \hpws for the different personas.
        For the vast majority of personas, the persona visiting $W^R$ receives more cookie synchronizations than the one visiting $W^L$.
        (b) Distribution of prices in CPM paid by advertisers to deliver ads to personas.
        The median price for $W^L$ is \leftPrices CPM, and the median price for $W^R$ is \rightPrices CPM.
        Considering the top 25\% most expensive ads, $W^R$ got (up to \diffPrices) higher ad-prices than $W^L$.
        (c) Heat-map with the KS statistic of 2-sample KS test for all pairwise comparisons between distributions of paid prices per persona, and visited $W^L$ and $W^R$.
        All cells with $p\geq 0.01$ are whited out; only cells with $p<0.01$ are colored.}
\end{figure*}

The results of this clustering have two implications.
First, the well-defined clusters for $k=3$ validate our hypothesis that the trackers deployed on \hpws preferentially converge onto specific partisan personas.
Second, it allows us to further pinpoint how left- and right-leaning personas are associated with different kinds of third parties, ranging from advertising to analytics. 
The value of $k=3$ also matches the intuition that two of the three clusters could be assigned to the left and right partisan positions, respectively.
The third cluster can be interpreted as the middle ground (if any) between party lines.
This can be seen in Figure~\ref{fig:nmfW}.
Figure~\ref{fig:nmfW}a (first party) shows that our profiles clearly fall into either the Left or Right cluster.
Figure~\ref{fig:nmfW}b (first party) shows only the top 10 websites for clarity (rather than all \hpws).
Here again, we can see websites fall into either the left or right category. 
After we computed the NMF on the different matrices $A^t$, we get the various $B^t$ and $C^t$ factors in Figure~\ref{fig:nmfW}, for advertising, analytics, etc.
Figure~\ref{fig:nmfW}a shows that `first party' and `other' domains essentially take clearly the extreme far clusters, as they have in cookies set from $W^L$ and $W^R$ (Figure~\ref{fig:nmfW}b).
Personas exposed to $W^L$ are easier to cluster for advertising, analytics and content, whereas social is less clear.
Furthermore, from the $C^t$ factors in Figure~\ref{fig:nmfW}b, we see that some first party domains are clearly clustered in the left (e.g., \emph{The Nation}) and others as right (e.g., \emph{Fox News}).
Also, for analytics and advertising, many domains are heavy on the right-wing cluster.
For content, many domains including \emph{google, microsoft} and \emph{adobe} appear to be clustered on the left-wing cluster, apart from \emph{oberon}\footnote{oberon-media.com is a multi-platform firm that gives gaming solutions}.
Interestingly, for social, \emph{Facebook} and \emph{reddit} are clustered in the right-wing cluster.

\subsection{User information flows and Cookie Synchronization between trackers}\label{sec:cs}

So far, we studied individual cookies set for each user (persona) in our methodology, and their potential impact on tracking of the said user.
In this section, we set out to explore the possible discrepancies in the flows of the user information among the different tracking and advertising entities.
The scope of these information flows is not only to share directly information about the current user, but more importantly to synchronize the different aliases by which a user may be known to different entities, building this way a common identifier.
This identifier will later allow the server-to-server exchange of user data in the background~\cite{acar2014web}, or data purchase in user data markets~\cite{Spiekermann2015, elmeleegy2013overview}.
This alias synchronization is the point where all user data collected by different third party trackers get attributed to a very single rich user profile.

The most common alias (or user ID) used on the web for user identification is the cookie ID, set on the user side the first time a domain meets the said user.
However, cookies are domain-specific.
So according to the same-origin policy, a domain cannot see the cookies set by another domain.
To bypass this and join together the different sets of user IDs, trackers and advertisers invented the \emph{Cookie Synchronization} (CS) technique, which they use in order to pass information from one to the other tagged with the ID of the user, thereby merging their databases in the background.

CS is performed when a third party $D1$, embedded in website $W$, redirects or issues a new HTTP request to a different collaborating third party $D2$, while piggybacking in the request the cookie ID $userD1$ it has set for the given user.
This way, $D2$ learns that the user just visited $W$ (information obtained from the referrer field) and is also known on the web as $user~D1$.
After that, $D2$ can drop its own cookie on the user side or sync a previously set cookie.
It is apparent that the more CS appear in a website, the more third parties can learn about the user.

To explore if there are any discrepancies in this ID synching and merging for the different categories of personas, we leverage 
existing CS detection methodologies~\cite{acar2014web,panpap_www2019}.
We apply these detection methodologies in our dataset, and in Figure~\ref{fig:csyncBox}, we plot for each crawl the detected CS per \hpw for the different personas. 
Although the ratio of CS per website varies per persona, for the vast majority of personas, the websites in $W^R$ not only set more cookies (as shown in Figure~\ref{fig:jsCookiesCDF}) but also perform more CSs than $W^L$.
By facilitating more CSs, $W^R$ enable more third party trackers to access and track the browsing behavior of the users than $W^L$.
This leads to more intense personalization and targeting of ads, which can be measured by the prices paid by advertisers for said ads.
We explore this in the next paragraphs.

\subsection{Which user costs more to advertisers?}\label{sec:prices}

Digital advertising is where user tracking and collected user data get translated to money.
Users get targeted with ads that match their interests, as extracted from the collected user data.  
As a next step, we set out to explore the monetary cost advertisers pay for the different personas, with the dimension of hyper-partisanship.
To achieve this, we leverage the most prevalent mechanism of digital ad-buying and delivery nowadays, i.e., the real-time programmatic ad-auctions~\cite{programmaticAd}.
Under these setting, available ad-slots in a publisher's website can be auctioned as soon as the website is rendered in the user's display.
This Vickrey type~\cite{vickrey1961counterspeculation} of auction takes place at real time in ad-markets called ad-exchanges (ADX) and special bidding engines called Demand-Side Platforms (DSPs).
These DSPs represent ad-agencies that place bids per auctioned ad-slot.
In order for the DSPs to decide if and how much they will bid for an ad-slot, detailed information is needed regarding the interests, geolocation, gender, age, sexual or political preferences, etc., of the current user viewing the webpage.
Such information is being bought from trackers and data warehouses on the background in an asynchronous fashion. 
To reduce latency, along with the bid, DSPs also send the ad-impression they want to deliver to the users if their bid wins the auction.

In order to assess the winning bid and thus the price paid for the ad-slots each user receives, we leverage the most dominant ad-auction protocol nowadays: the RTB protocol~\cite{RTB}. 
Specifically, we capture the step where the user's browser (as instructed by the ADX) issues a HTTP request to the winning bidder notifying them about: (i) its win, (ii) the charge price, and (iii) the successful rendering of the ad-impression.
Following the methodology of previous works~\cite{papadopoulos2017if,castelluccia2014selling}, we estimate the monetary values the ad-ecosystem paid for the different category of personas in our dataset.

In Figure~\ref{fig:pricesCDF}, we present the distributions of the ad-prices for $W^L$ and $W^R$.
As we can observe, the median price for $W^L$ is \leftPrices CPM (CPM=Cost Per Mille, i.e., cost per thousand impressions), and the median price for $W^R$ is \rightPrices CPM.
However, when it comes to the top quartile of the more expensive ads, $W^R$ get (up to \diffPrices) higher prices than the corresponding ones of $W^L$.

Finally, to explore the variability of prices across different personas, in Figure~\ref{fig:pricesKS}, we plot a heat-map with the statistics of 2-sample KS tests of all pairwise comparisons between the paid ad-prices per persona, for $W^L$ and $W^R$.
We not only see a particular variability among the different types of \hpws as demonstrated in Figure~\ref{fig:pricesCDF}, but we also see that there are personas with significantly higher ad-prices than others for the same type of websites.

\subsection{Summary}
In general, our results indicate that right-leaning \hpws have more third parties embedded in their pages, they monitor and track users with more cookies, these third parties synchronize their cookies more often or with more intensity, and all this tracking activity leads to better ad-targeting, and thus higher paid prices for delivered ads.
We summarize our findings as follows:

\begin{enumerate}
    \item Right-leaning \hpws news have embedded significantly more third parties and store up to 25\% more cookies to user browsers than left-leaning \hpws.
    \item User personas exhibiting realistic and representative demographic characteristics tend to receive up to 15\% more tracking (cookies) from \hpws than baseline personas (i.e., with no characteristics).
    \item Popular, highly-ranked \hpws track users more intensely than lower-ranked \hpws.
    \item Single-feature personas, e.g., Woman, Man, Youth, are highly tracked by default, no matter what party-leaning they demonstrate through their visits.
    \item It is possible to group personas to left or right-leaning clusters, based on the domains and numbers of cookies they receive while browsing \hpws.
    \item Right-leaning \hpws aid third parties to track more intensely their users than left-leaning \hpws, by facilitating up to 50\% more cookie synchronizations between trackers than left-leaning \hpws.
    \item Right-leaning \hpws deliver to users ads costing up to 5x more than in left-leaning \hpws, as a consequence of more intense tracking and ad-targeting performed.
\end{enumerate}

\section{Related Work}
\label{Sec:related}

\noindent{\bf Politics and Polarization.}
Political parties involve user audience often through polarization, as well as by urging them to volunteer and campaign for the party~\cite{ballard2016campaigning}.
With most of the political discussion happening in the cyber-space, user experience leads the way of engagement and enchantment~\cite{mccarthy2006experience, agarwal2019tweeting}.
For example, news personalization is a way to send curated news briefing to subscribers~\cite{messing2014selective,thurman2012future}.
\cite{chand2017dark} coined dark money which is spent on negative ads.
Another selective political initiative is the highly polarized \hpws in US politics, studied by~\cite{bhatt2018illuminating}.
\hpws declare themselves right- or left-leaning in their web portals.
Interestingly, many of these \hpws were deleted after US 2016 elections. 
However, more than 550 still exist and we focus our study on them.
Alternate (or fake) news websites can also be highly engaging.
URL sharing of alternate news across the Web was studied in~\cite{zannettou2017web}.
Studies have tried to understand the political stance of users and news websites by various methods.
Examples include the use of NMF clustering ($k=2$)~\cite{lahoti2018joint}, finding non-binary political ideological scores, sentiment analysis~\cite{pla2014political}, twitter profile data~\cite{pennacchiotti2011democrats}, etc.

\noindent{\bf User Tracking and Information flow.}
Many studies have analyzed the extent of online user tracking.
In~\cite{karaj2018whotracks}, authors measure the trackers per website and find that, on average, there are 10 different trackers per website; in news websites this average is 15 trackers.
Similarly, in~\cite{englehardt2016online}, authors used Alexa top 500 websites for each of the 16 basic content categories of Alexa.
They find that websites providing editorial content embed the largest amount of trackers given their lack of external funding sources, and their need to monetize page views through advertising.

A popular web tracking mechanism is the use of cookies set on the user-side.
In~\cite{hu2019characterising}, researchers measured differences in third party cookies offloaded on a browser by collecting data from real users over a period of 1 year, before and after the GDPR regulation~\cite{europeanDataRules2018}, and found that the number of cookies does not change, not least because users tend to pay little attention to the GDPR options.
In~\cite{castelluccia2014selling}, researchers were the first to analyze the technique of synchronizing different user IDs that various entities set for the same user, thus reducing the user's anonymity on the Web.
They also study Cookie Synchronization (CS) in conjunction with the ad-delivery paradigm of RTB auctions.
In~\cite{costOfAds}, researchers measure the costs of digital advertising and use CS as a metric for the reduction of user's privacy.
Also,~\cite{panpap_www2019} conducted a thorough analysis of CS by analyzing a year-long dataset of real mobile users and showed that CS increases the number of domains that track a user's browsing by a factor of 6.75.
In~\cite{exclusiveCSync} authors show that CS can break users' secure SSL sessions and increase their identifiability on the Web, even if they browse over VPN.
They probed 12,000 top Alexa websites and found that 1 out of 13 websites expose their visitors to this severe privacy leak.
In~\cite{acar2014web}, authors study CS together with re-spawning cookies and investigate how these two can form persistent cookies that users cannot abolish even after clearing their cookie jar.
In~\cite{englehardt2015cookies}, authors simulate users browsing the web in an attempt to study the possibility of a passive eavesdropper leveraging third party HTTP tracking cookies for mass monitoring.
They find that indeed the eavesdropper can reconstruct 62-73\% of a typical user's browsing history.

\noindent{\bf Biases in Online Ecosystem.}
In an attempt to shed light over today's opaque online ad-ecosystem,~\cite{lecuyer2015sunlight} proposed Sunlight: a system to detect targeting in large-scale experiments with statistically justifiable results.
Their system consists of four components that generate, refine, and validate hypotheses to reveal the possible causes of observed targeting.
Similarly,~\cite{bashir2016} developed a content-agnostic methodology to detect client- and server-side flows of information between ad-exchanges by leveraging retargeted ads.
In~\cite{zhang2017targeted}, user browsing history is studied to define audience gender and age demographics.
An SVM model is trained on the text in websites, to predict user features and bias in advertising.
Evidences of bias in ad-placement is given by~\cite{datta2015automated}.
Using a tool named AdFisher, the authors investigate ad-placement with adjustment of Google ads settings of user.
Thus, setting gender to female resulted in fewer ads related to high paying jobs.
In~\cite{kim2018stealth}, authors studied 5 million paid ads on Facebook and claim that US users are mostly targeted by foreign groups using loopholes in regulations of Facebook.

\section{Discussion and Future Work} \label{sec:conclusion}

Understanding the dissemination of misinformation over online platforms and fringe news sources (e.g., Breitbart or Infowars) has become a crucial topic of study because of its considerable impact on our online culture.
These sources are often found responsible for spreading news and opinion pieces with extreme views along partisan lines.
This strategy keeps users engaged with the websites for prolonged periods of time.
A major incentive of such \hpwlongs (\hpws) to follow this strategy is monetization by delivering targeted and personalized ad-content to profiled users.
In fact, these websites do not need to provide any credible content or news, thus, reducing costs for journalists and other expenses, otherwise incurred on legitimate news websites.
However, until now, it remained unclear how such \hpws perform targeted advertising and whether they \emph{differentiate} their online user tracking based on demographics and partisan leaning of their audience.

To shed light to this problem, in this paper, we present a methodology that creates artificial users (personas) with certain demographic features.
We find that a stateful browsing of a small number of 10 websites is sufficient to build a persona (cookie state) of hundreds of unique third parties.
We store browser states of personas and use them to emulate diverse types of browsing patterns along partisan lines, and in the process, record how these websites track them.
The data acquired from such persona-based crawls are an asset for investigating user experience, bias in tracking, and various other privacy metrics.
By visiting \visitedSites \hpws, we collected and analyzed a dataset of $\sim$19 million HTTP (requests, responses and redirects) calls and $\sim$3.5 million cookies.
In general, we find a higher amount of tracking in the right-leaning \hpws, and also an increased tracking for personas exhibiting specific patterns with respect to a combination of gender and age demographics.

Among the right-leaning \hpws, the top ranked (by Alexa) websites are consistently loaded with more trackers, in comparison to less popular \hpws.
By performing a co-clustering on the personas using the NMF model, we observe that we can group personas and third parties in a low dimensional space using the types of cookie domains as features.
We obtain three optimal clusters which can be labelled as (mild) left, (mild) right and neutral.
Finally, investigating the third party cookie synchronization (CS) and examining the prices paid for ads as declared in the real-time bidding protocol, we conclude that right-leaning \hpws apply more intense CS and command higher prices for delivered ads.

For the benefit of reproducibility, we reported an in-depth process for the creation and deployment of said personas.
This approach is easily scalable across different demographic features and browsing patterns.
Our framework can also be repurposed for auditing other ecosystems such as  e-commerce, web search engines, public social media pages, and other recommendation services.
Our framework can be useful for enforcing new regulations under data protection (e.g., GDPR~\cite{europeanDataRules2018}) which have been introduced to give back control of personal data to their owners, and increase overall Web transparency.
As GDPR enforcement is in its initial phase, most of the control on user's data and tracking still remains with the websites, allowing many \hpws to take liberties on how to comply with regulations.
For example, in our study, we found cases of right-leaning \hpws (e.g., \url{spectator.org}) and left-leaning \hpws (e.g., \url{newshounds.us}) setting more than 1000 cookies per user.

Indeed, there are some limitations of our framework, which we plan to overcome in the near future.
For example, data protection laws do not allow cross-border tracking, which results in logging 33\% fewer trackers when the crawling is performed from websites under EU's jurisdiction, as compared to traffic logged within US.
Another area to explore is the variability in tracking of personas, and its relation with the ad-economies and end-user experience.
For example, if a website is visited on two consecutive days, tracking metrics will most likely be different.
Also, there is a wide-range of different, less or more pronounced and compound personas that someone can emulate with our framework.
Clearly, every persona comes with costs of finding appropriate websites to build it, and additional experiments to make sure it triggers network traffic realistic for an online user of that persona.
In fact, less pronounced personas may not be able to collect appropriate and conclusive data of differential tracking.
In the future, we aim to build better tools to understand these problems.
We also plan to further develop our framework to look into the ad-ecosystem on fringe news websites, and replicate our work for multiple countries to discover user differential tracking and personal data leaks.

\section{Acknowledgements}

The research leading to these results has received funding from the EU's Horizon 2020 Programme under grant agreements No 830927 (project CONCORDIA) and No 871370 (project PIMCITY). We also acknowledge  support via  EPSRC Grant Ref: EP/T001569/1 for ``Artificial Intelligence for Science, Engineering, Health and Government'', and particularly the ``Tools, Practices and Systems'' theme for ``Detecting and Understanding Harmful Content Online: A Metatool Approach'', as well as a King's India Scholarship 2015 and a Professor Sir Richard Trainor Scholarship 2017 at King's College London.
The paper reflects only the authors' views and the Agency and the Commission are not responsible for any use that may be made of the information it contains.

\bibliographystyle{unsrt}  

\bibliography{bibliography}

\begin{thebibliography}{10}

\bibitem{costOfAds}
Panagiotis Papadopoulos, Nicolas Kourtellis, and Evangelos~P. Markatos.
\newblock The cost of digital advertisement: Comparing user and advertiser
  views.
\newblock In {\em Proceedings of the World Wide Web Conference}, pages
  1479--1489, 2018.

\bibitem{panpap_www2019}
Panagiotis Papadopoulos, Nicolas Kourtellis, and Evangelos~P. Markatos.
\newblock Cookie synchronization: Everything you always wanted to know but were
  afraid to ask.
\newblock In {\em Proceedings of the 28th International Conference on World
  Wide Web}, 2019.

\bibitem{papadopoulos2017if}
Panagiotis Papadopoulos, Nicolas Kourtellis, Pablo~Rodriguez Rodriguez, and
  Nikolaos Laoutaris.
\newblock If you are not paying for it, you are the product: How much do
  advertisers pay to reach you?
\newblock In {\em Proceedings of the Internet Measurement Conference}, pages
  142--156. ACM, 2017.

\bibitem{pachilakis2019headerbidding}
Michalis Pachilakis, Panagiotis Papadopoulos, Evangelos~P. Markatos, and
  Nicolas Kourtellis.
\newblock {No More Chasing Waterfalls: A Measurement Study of the Header
  Bidding Ad-Ecosystem}.
\newblock In {\em Proceedings of the Internet Measurement Conference}, pages
  280--293. ACM, 2019.

\bibitem{castelluccia2014selling}
Lukasz Olejnik, Tran Minh-Dung, and Claude Castelluccia.
\newblock Selling off privacy at auction.
\newblock In {\em Network and Distributed System Security Symposium (NDSS)},
  2014.

\bibitem{acar2014web}
Gunes Acar, Christian Eubank, Steven Englehardt, Marc Juarez, Arvind Narayanan,
  and Claudia Diaz.
\newblock The web never forgets: Persistent tracking mechanisms in the wild.
\newblock In {\em Proceedings of the SIGSAC Conference on Computer and
  Communications Security}, pages 674--689. ACM, 2014.

\bibitem{Nikiforakis:2013:CME:2497621.2498133}
Nick Nikiforakis, Alexandros Kapravelos, Wouter Joosen, Christopher Kruegel,
  Frank Piessens, and Giovanni Vigna.
\newblock Cookieless monster: Exploring the ecosystem of web-based device
  fingerprinting.
\newblock In {\em Proceedings of the Symposium on Security and Privacy}, pages
  541--555. IEEE Computer Society, 2013.

\bibitem{devFingerprinting}
Elias~P. Papadopoulos, Michalis Diamantaris, Panagiotis Papadopoulos, Thanasis
  Petsas, Sotiris Ioannidis, and Evangelos~P. Markatos.
\newblock The long-standing privacy debate: Mobile websites vs mobile apps.
\newblock In {\em Proceedings of the 26th International Conference on World
  Wide Web}, pages 153--162, 2017.

\bibitem{CATrump}
Paul~Hilder Paul~Lewis.
\newblock Leaked: Cambridge analytica's blueprint for trump victory.
\newblock
  \url{https://www.theguardian.com/uk-news/2018/mar/23/leaked-cambridge-analyticas-blueprint-for-trump-victory},
  2018.

\bibitem{CACruz}
Patrick Svitek and Haley Samsel.
\newblock Ted cruz says cambridge analytica told his presidential campaign its
  data use was legal.
\newblock
  \url{https://www.texastribune.org/2018/03/20/ted-cruz-campaign-cambridge-analytica/},
  2018.

\bibitem{CABrexit}
Peter Geoghegan and Jenna Corderoy.
\newblock Revealed: Arron banks brexit campaign's secret meetings with
  cambridge analytica.
\newblock
  \url{https://www.opendemocracy.net/uk/brexitinc/peter-geoghegan-jenna-corderoy/revealed-arron-banks-brexit-campaign-had-more-meetings-w},
  2018.

\bibitem{RussiaFB}
Leonid Bershidsky.
\newblock Yes, russia abused facebook. but did it work?
\newblock
  \url{https://www.bloomberg.com/opinion/articles/2018-12-18/yes-russia-abused-facebook-but-did-it-work},
  2018.

\bibitem{bhatt2018illuminating}
Shweta Bhatt, Sagar Joglekar, Shehar Bano, and Nishanth Sastry.
\newblock Illuminating an ecosystem of partisan websites.
\newblock In {\em {Companion of The Web Conference}}, pages 545--554, 2018.

\bibitem{lee1999learning}
Daniel~D Lee and H~Sebastian Seung.
\newblock Learning the parts of objects by non-negative matrix factorization.
\newblock {\em Nature}, 401(6755):788, 1999.

\bibitem{lahoti2018joint}
Preethi Lahoti, Kiran Garimella, and Aristides Gionis.
\newblock Joint non-negative matrix factorization for learning ideological
  leaning on twitter.
\newblock In {\em Proceedings of the 11th International Conference on Web
  Search and Data Mining}, pages 351--359. ACM, 2018.

\bibitem{Zhu:2007:CCL:1277741.1277825}
Shenghuo Zhu, Kai Yu, Yun Chi, and Yihong Gong.
\newblock Combining content and link for classification using matrix
  factorization.
\newblock In {\em Proceedings of the 30th SIGIR Conference on Research and
  Development in Information Retrieval}, pages 487--494. ACM, 2007.

\bibitem{itp}
John Wilander.
\newblock Intelligent tracking prevention 2.3.
\newblock
  \url{https://webkit.org/blog/9521/intelligent-tracking-prevention-2-3/},
  2019.

\bibitem{sop}
{World Wide Web Consortium (W3C)}.
\newblock Same origin policy, 2010.

\bibitem{bashir2016}
Muhammad~Ahmad Bashir, Sajjad Arshad, William Robertson, and Christo Wilson.
\newblock Tracing information flows between ad exchanges using retargeted ads.
\newblock In {\em {25th USENIX Security Symposium}}, 2016.

\bibitem{harding2001cookies}
William~T Harding, Anita~J Reed, and Robert~L Gray.
\newblock Cookies and web bugs: What they are and how they work together.
\newblock {\em Information Systems Management}, 18(3):17--24, 2001.

\bibitem{solomos2019cdt}
Konstantinos Solomos, Panagiotis Ilia, Sotiris Ioannidis, and Nicolas
  Kourtellis.
\newblock {Talon: An Automated Framework for Cross-Device Tracking Detection}.
\newblock In {\em International Symposium on Research in Attacks, Intrusions
  and Defenses (RAID)}, pages 227--241, 2019.

\bibitem{jernigan2009gaydar}
Carter Jernigan and Behram~FT Mistree.
\newblock Gaydar: Facebook friendships expose sexual orientation.
\newblock {\em First Monday}, 14(10), 2009.

\bibitem{toysmart}
{The Federal Trade Commission}.
\newblock Ftc sues failed website, toysmart.com, for deceptively offering for
  sale personal information of website visitors.
\newblock
  \url{https://www.ftc.gov/news-events/press-releases/2000/07/ftc-sues-failed-website-toysmartcom-deceptively-offering-sale},
  2000.

\bibitem{nsaCookies}
Seth~Schoen Adi~Kamdar, Rainey~Reitman.
\newblock Nsa turns cookies (and more) into surveillance beacons.
\newblock
  \url{https://www.eff.org/deeplinks/2013/12/nsa-turns-cookies-and-more-surveillance-beacons},
  2013.

\bibitem{Buzzfeed}
Criag Silverman, Jane Lytvynenko, Lam Thuy~Vo, and Jeremy Singer-Vine.
\newblock Inside the partisan fight for your news feed, 2017.
\newblock
  https://www.buzzfeednews.com/article/craigsilverman/inside-the-partisan-fight-for-your-news-feed.

\bibitem{BuzzfeedGithub}
Criag Silverman, Jane Lytvynenko, Lam Thuy~Vo, and Jeremy Singer-Vine.
\newblock Data, analytic code, and findings related to the buzzfeed news
  article, 2017.
\newblock
  https://github.com/BuzzFeedNews/2017-08-partisan-sites-and-facebook-pages.

\bibitem{carrascosa2015}
Juan~Miguel Carrascosa, Jakub Mikians, Ruben Cuevas, Vijay Erramilli, and
  Nikolaos Laoutaris.
\newblock I always feel like somebody's watching me: Measuring online
  behavioural advertising.
\newblock In {\em Proceedings of the 11th ACM Conference on Emerging Networking
  Experiments and Technologies}, CoNEXT '15, 2015.

\bibitem{programmaticAd}
{Adext Corp}.
\newblock How programmatic advertising has changed the online advertising
  landscape.
\newblock https://blog.adext.com/programmatic-ads-changed-online-advertising/,
  2019.

\bibitem{ballard2016campaigning}
Andrew~O Ballard, D~Sunshine Hillygus, and Tobias Konitzer.
\newblock Campaigning online: Web display ads in the 2012 presidential
  campaign.
\newblock {\em PS: Political Science \& Politics}, 49(3):414--419, 2016.

\bibitem{englehardt2016online}
Steven Englehardt and Arvind Narayanan.
\newblock Online tracking: A 1-million-site measurement and analysis.
\newblock In {\em Proceedings of the SIGSAC Conference on Computer and
  Communications Security}, pages 1388--1401. ACM, 2016.

\bibitem{coppa}
Federal~Trade Commission.
\newblock Children's online privacy protection rule (coppa), 1998.
\newblock
  https://www.ftc.gov/enforcement/rules/rulemaking-regulatory-reform-proceedings/childrens-online-privacy-protection-rule.

\bibitem{lecuyer2015sunlight}
Mathias Lecuyer, Riley Spahn, Yannis Spiliopolous, Augustin Chaintreau, Roxana
  Geambasu, and Daniel Hsu.
\newblock Sunlight: Fine-grained targeting detection at scale with statistical
  confidence.
\newblock In {\em Proceedings of the 22nd SIGSAC Conference on Computer and
  Communications Security}, pages 554--566. ACM, 2015.

\bibitem{openwpm-code}
Openwpm framework.
\newblock \url{https://github.com/mozilla/OpenWPM}, 2020.

\bibitem{disconnectme}
Disconnect.me.
\newblock Disconnect tracking protection, 2019.
\newblock
  https://github.com/disconnectme/disconnect-tracking-protection/blob/master/services.json.

\bibitem{whotracksme}
whotracks.me~by Cliqz and Ghostery.
\newblock Bringing transparency to online tracking., 2019.
\newblock
  https://github.com/cliqz-oss/whotracks.me/tree/master/whotracksme/data/assets.

\bibitem{brunet2004metagenes}
Jean-Philippe Brunet, Pablo Tamayo, Todd~R Golub, and Jill~P Mesirov.
\newblock Metagenes and molecular pattern discovery using matrix factorization.
\newblock {\em Proceedings of the National Academy of Sciences},
  101(12):4164--4169, 2004.

\bibitem{hutchins2008position}
Lucie~N Hutchins, Sean~M Murphy, Priyam Singh, and Joel~H Graber.
\newblock Position-dependent motif characterization using non-negative matrix
  factorization.
\newblock {\em Bioinformatics}, 24(23):2684--2690, 2008.

\bibitem{Spiekermann2015}
Sarah Spiekermann, Rainer B{\"o}hme, Alessandro Acquisti, and Kai-Lung Hui.
\newblock Personal data markets.
\newblock {\em Electronic Markets}, 25(2):91--93, Jun 2015.

\bibitem{elmeleegy2013overview}
Hazem Elmeleegy, Yinan Li, Yan Qi, Peter Wilmot, Mingxi Wu, Santanu Kolay, Ali
  Dasdan, and Songting Chen.
\newblock Overview of turn data management platform for digital advertising.
\newblock {\em Proceedings of the VLDB Endowment}, 6(11):1138--1149, 2013.

\bibitem{vickrey1961counterspeculation}
William Vickrey.
\newblock Counterspeculation, auctions, and competitive sealed tenders.
\newblock {\em The Journal of finance}, 1961.

\bibitem{RTB}
Jack Marshall.
\newblock What is real time bidding.
\newblock https://digiday.com/media/what-is-real-time-bidding/, 2014.

\bibitem{mccarthy2006experience}
John McCarthy, Peter Wright, Jayne Wallace, and Andy Dearden.
\newblock The experience of enchantment in human--computer interaction.
\newblock {\em Personal and ubiquitous computing}, 10(6):369--378, 2006.

\bibitem{agarwal2019tweeting}
Pushkal Agarwal, Nishanth Sastry, and Edward Wood.
\newblock Tweeting mps: Digital engagement between citizens and members of
  parliament in the uk.
\newblock In {\em Proceedings of the International AAAI Conference on Web and
  Social Media}, volume~13, pages 26--37, 2019.

\bibitem{messing2014selective}
Solomon Messing and Sean~J Westwood.
\newblock Selective exposure in the age of social media: Endorsements trump
  partisan source affiliation when selecting news online.
\newblock {\em Communication Research}, 41(8):1042--1063, 2014.

\bibitem{thurman2012future}
Neil Thurman and Steve Schifferes.
\newblock The future of personalization at news websites: lessons from a
  longitudinal study.
\newblock {\em Journalism Studies}, 13(5-6):775--790, 2012.

\bibitem{chand2017dark}
Daniel~E Chand.
\newblock Dark money and dirty politics: Are anonymous ads more negative?
\newblock {\em Business and Politics}, 19(3):454--481, 2017.

\bibitem{zannettou2017web}
Savvas Zannettou, Tristan Caulfield, Emiliano De~Cristofaro, Nicolas
  Kourtellis, Ilias Leontiadis, Michael Sirivianos, Gianluca Stringhini, and
  Jeremy Blackburn.
\newblock The web centipede: understanding how web communities influence each
  other through the lens of mainstream and alternative news sources.
\newblock In {\em Proceedings of the 2017 Internet Measurement Conference},
  pages 405--417. ACM, 2017.

\bibitem{pla2014political}
Ferran Pla and Llu{\'\i}s-F Hurtado.
\newblock Political tendency identification in twitter using sentiment analysis
  techniques.
\newblock In {\em Proceedings of COLING 2014, the 25th international conference
  on computational linguistics: Technical Papers}, pages 183--192, 2014.

\bibitem{pennacchiotti2011democrats}
Marco Pennacchiotti and Ana-Maria Popescu.
\newblock Democrats, republicans and starbucks afficionados: user
  classification in twitter.
\newblock In {\em Proceedings of the 17th SIGKDD International Conference on
  Knowledge Discovery and Data mining}, pages 430--438. ACM, 2011.

\bibitem{karaj2018whotracks}
Arjaldo Karaj, Sam Macbeth, R{\'e}mi Berson, and Josep~M Pujol.
\newblock Whotracks.me: Monitoring the online tracking landscape at scale.
\newblock {\em arXiv preprint arXiv:1804.08959}, 2018.

\bibitem{hu2019characterising}
Xuehui Hu and Nishanth Sastry.
\newblock Characterising third party cookie usage in the eu after gdpr.
\newblock In {\em Proceedings of the 10th ACM Conference on Web Science}, pages
  137--141, 2019.

\bibitem{europeanDataRules2018}
{European Commission}.
\newblock Rules for the protection of personal data inside and outside the eu.
\newblock https://ec.europa.eu/info/law/law-topic/data-protection\_en/, 2018.

\bibitem{exclusiveCSync}
Panagiotis Papadopoulos, Nicolas Kourtellis, and Evangelos~P. Markatos.
\newblock Exclusive: How the (synced) cookie monster breached my encrypted vpn
  session.
\newblock In {\em Proceedings of the 11th European Workshop on Systems
  Security}, EuroSec'18, pages 6:1--6:6, 2018.

\bibitem{englehardt2015cookies}
Steven Englehardt, Dillon Reisman, Christian Eubank, Peter Zimmerman, Jonathan
  Mayer, Arvind Narayanan, and Edward~W Felten.
\newblock Cookies that give you away: The surveillance implications of web
  tracking.
\newblock In {\em Proceedings of the 24th International Conference on World
  Wide Web}, pages 289--299, 2015.

\bibitem{zhang2017targeted}
Yong Zhang, Hongming Zhou, Nganmeng Tan, Saeed Bagheri, and Meng~Joo Er.
\newblock Targeted advertising based on browsing history.
\newblock {\em arXiv preprint arXiv:1711.04498}, 2017.

\bibitem{datta2015automated}
Amit Datta, Michael~Carl Tschantz, and Anupam Datta.
\newblock Automated experiments on ad privacy settings.
\newblock {\em Proceedings on Privacy Enhancing Technologies}, 2015(1):92--112,
  2015.

\bibitem{kim2018stealth}
Young~Mie Kim, Jordan Hsu, David Neiman, Colin Kou, Levi Bankston, Soo~Yun Kim,
  Richard Heinrich, Robyn Baragwanath, and Garvesh Raskutti.
\newblock The stealth media? groups and targets behind divisive issue campaigns
  on facebook.
\newblock {\em Political Communication}, pages 1--27, 2018.

\end{thebibliography}

\end{document}